\documentclass[pre,aps,nofootinbib,showpacs,showkeys]{revtex4}
\usepackage{graphicx}
\usepackage{natbib}
\usepackage{bm}
\usepackage{amssymb}
\usepackage{amsmath}
\usepackage{euscript}
\usepackage{esint}
\usepackage{slashed}

\begin{document}

\title{Analytical study in the mechanism of flame movement in horizontal tubes}

\author{Kirill A. Kazakov}
\affiliation{%
Department of Theoretical Physics, Physics Faculty, Moscow State
University, 119991, Moscow, Russian Federation}

\begin{abstract}
The problem of premixed flame propagation in wide horizontal tubes is revisited. Employing the on-shell description of flames with arbitrary gas expansion, a nonlinear second-order differential equation for the front position of steady flame is derived. Solutions to this equation, obtained numerically, reveal two distinct physical regimes of laminar flame propagation controlled by the strong baroclinic effect. They differ by the front shape and flame speed, the ratio of the total consumption rates in the two regimes being $1.4$ to $1.8,$ depending on the value of the gas expansion coefficient. Comparison with the existing experimental data on methane-air flames is made, and explanation of the main trends in the observed flame behavior is given. It is shown, in particular, that the faster (slower) regime of combustion is realized in mixtures close to (far from) the stoichiometric composition, with pronounced changeover in between.
\end{abstract}
\maketitle

\section{Introduction}

All laboratory flames are subject to the terrestrial gravitational field whose influence on the process of deflagration is often prominent. In fact, for many premixed flames evolving spontaneously this influence is dominant as regards formation of the large-scale flame structure. This is because deflagrations involve flows of markedly different densities, characterized by essentially subsonic fluid velocities. A simple dimensional analysis suggests that under these conditions, the relative gravity impact can be estimated as $(acceleration~of~gravity)\times  (characteristic~length)/(characteristic~velocity)^2.$ It follows that propagation of all hydrocarbon-air flames in tubes wider than 1\,cm is strongly affected by the gravitational field, the impact being especially pronounced for flames with low local consumption rates. Observations fully confirm this conclusion. Under the action of gravity, the front area considerably increases, and proportionally increases the flame speed. In the regime of uniform movement, {\it i.e.,} when the flame speed remains grossly constant, it may be tens times the normal flame speed. For instance, the normal flame speed in the mixture of $6\%$ methane with air under standard conditions is about 0.1\,m/s, while its propagation speed in tubes of 1\,m diameter is 1.5\,m/s.

A vulgar description of the gravitational influence on flame dynamics is that it is a kind of buoyancy effect, and in the case of flame propagation in horizontal tubes goes as follows \cite{zeldo1985}. The light products of combustion moving upward tend to occupy space above the heavy fresh gas, which causes the flame front to tilt and spread along the tube. This view is actually inadequate, if only because it implies that the process of flame spreading ought to last indefinitely. Indeed, if buoyancy were the driving force, any tilted flame configuration would be unstable against further front flattening, which would result in unbounded growth of the flame speed. As a matter of fact, no description of flame behavior in terms of forces can be adequate, because this behavior is essentially determined by the properties of the mass flow through the flame front, and it is flow velocities that are the right terms to describe these properties. The true mechanism governing flame evolution in the strong gravitational field is the baroclinic effect -- the gravity-induced generation of vorticity inside the flame front. It is thus vortical motion of the burnt gases that is responsible for the formation of titled flame configuration. Flame spreading stabilizes when vorticity generated by the flame front curvature becomes comparable to the baroclinic vorticity (Cf. Sec.~\ref{reduction}).

In view of practical and theoretical importance of horizontal flame propagation, it was thoroughly investigated already at the dawn of modern combustion science. The very existence of the regime of uniform movement was established by Mallard and Le Chatelier \cite{mallard1883}, after which it was scrutinized by Mason and Wheeler \cite{mason1917}, Mason \cite{mason1923}, Coward and Greenwald \cite{coward1928}, and in the famous works by Coward and Hartwell \cite{coward1932I,coward1932II}. The latter summarized and extended the results of the preceding works to infer normal flame speeds in mixtures of various composition, and to establish relationship between the propagation speed and tube diameter, which was shown to be more complicated than the simple power law.

Despite extensive experimental data acquired on the horizontal flame propagation, there has been lack of adequate theoretical description of this process. The reason for this is obvious: The large increase of flame velocity caused by the gravitational field implies that the process is strongly nonlinear, and hence is out of reach of the classic approach based on explicit solving of the bulk flow equations. The necessity to solve these equations explicitly is indeed the stumbling block of this approach already in the case of freely propagating flames, which restricts its applicability to flames with weak gas expansion. Namely, the fresh-to-burnt gas-density ratio, $\theta,$ must be close to unity in order to damp the nonlinear effects \cite{siv1977}. Under this condition, the horizontal flame propagation was considered in \cite{rakib1988}. The relative flame velocity increase within the small $(\theta-1)$-expansion is small by definition, as is vorticity produced by the flame. At the same time, laboratory flames have $\theta=5-10,$ and produce strongly nonlinear rotational flows of burnt matter. Therefore, application of the weak gas expansion approximation to flames in horizontal tubes can be only of academic interest. Furthermore, the above remark on the role of the baroclinic effect shows that the potential flow models such as Frankel's \cite{frankel1990}, or the bubble model \cite{bychkov1997} are also irrelevant. The latter, for instance, replaces the flame by an open inert bubble moving in a horizontal channel, and gives for the steady flame speed the value $0.43\sqrt{gb}$ ($g$ is the gravity acceleration, $b$ the channel width), which badly disagrees with observations. As was already mentioned, relation between the flame speed and tube diameter (channel width) is not that simple.

A way round the problem of solving the bulk equations was found in \cite{kazakov1,kazakov2} where exact equations for the fresh gas velocity at the flame front (the {\it on-shell} velocity, for brevity) were obtained for two-dimensional (2D) steady flame with arbitrary gas expansion. Specifically, it was shown that the velocity field of burnt gases can be decomposed into rotational and potential components in such a way that the on-shell value of the former can be explicitly expressed in terms of the on-shell fresh gas velocity, while retaining incompressibility and boundedness of the latter. These properties of the unknown potential component were then used to exclude it from the set of jump conditions at the front by applying an integral identity (the {\it dispersion relation}) following from the Cauchy theorem. The result is an exact complex integro-differential equation (the {\it master equation}) for the on-shell fresh gas velocity and the flame front position, which together with the evolution equation constitutes a closed system. This {\it on-shell formulation} was subsequently generalized to unsteady flames \cite{jerk1,jerk2}, and applied to various essentially nonperturbative problems \cite{kazakov3,jerk3,jerk4}.

The purpose of this paper is to give rigorous theoretical description of horizontal flame propagation. Using the on-shell formulation, it will be shown that the problem of uniform flame movement admits complete solution in the 2D case. Namely, the set of the on-shell equations will be reduced in Sec.~\ref{reduction} to a single second-order differential equation for the front position. This reduction relies on the large front-slope asymptotic expansion of the master equation, constructed in \cite{kazakov3}. Experiments show that in sufficiently wide tubes, flames are subject to a kind of turbulent motion observed as a swirling of the flame front. Occurrence of a small-scale structure generally makes the large-slope approximation inapplicable. An auxiliary but important result proved in Sec.~\ref{widechan} is that regarding the large-scale flame evolution, the leading-order effect of this structure is to renormalize the normal flame speed; the proof follows closely \cite{kazakov2003}. This theorem makes it possible to apply to wrinkled flames the results obtained for smooth patterns. Numerical solutions of the second-order equation are found in Secs.~\ref{numeric}, \ref{ident} where it is shown that for each set of physical parameters, there are two distinct regimes of laminar flame propagation, characterized by significantly different flame speeds. A detailed comparison of the obtained results with the experimental data on methane-air flames is carried out in Sec.~\ref{comparison}. The arguments justifying application of the 2D picture to real flames are summarized in Sec.~\ref{rel}. The effect of heat losses to the tube walls is discussed in Sec.~\ref{losses} where it is shown that because of the large flame spread along the tube, these losses affect horizontal flame propagation in tubes of any diameter. After assessment of the calculational accuracy, given in Sec.~\ref{assessment}, theoretical results are used in Sec.~\ref{comparison} to explain the measured flame speed versus tube diameter in mixtures of various methane concentration. It is demonstrated, in particular, that mixtures close to (far from) the stoichiometric composition favor the faster (slower) regime of flame propagation. Conclusions are drawn in Sec.~\ref{conclusions}.

\section{On-shell description of horizontally propagating flames}\label{onshell}

\subsection{Master equation}\label{mequation}

Consider a 2D flame propagating in an initially quiescent gaseous mixture filling horizontal channel of width $b$ with ideal walls. Relevance of this picture to the three-dimensional laboratory flames will be discussed in Sec.~\ref{rel}.
We choose Cartesian coordinates $(x_1,x_2) \equiv (x,y)$ so that the $y$-axis is along the upper channel wall, $y = - \infty$ being in the fresh gas. Then the flame front position at time $t$ can be described by an equation $y=f(x,t).$ Let $(v_1,v_2)\equiv (w,u)$ denote Cartesian components of the gas velocity vector $\bm{v}.$ Even though the flame speed in horizontal tubes often considerably exceeds the normal speed, gas flows induced by the propagating flame remain essentially subsonic. This allows the fluid be considered as incompressible, and its density, $\rho,$ be taken constant on both sides of the flame front. The fresh gas will be assumed throughout to be of unit density, so that the burnt gas density is $1/\theta.$ Also, it is often convenient to use natural units in which $b=U_f=1,$ where $U_f$ is the planar flame speed relative to the fresh gas (see also Sec.~\ref{renorm}).

Below, we will be primarily concerned with the steady regime of flame propagation, which is closely related to the notion of uniform flame movement. It will be understood that the steady system is considered in the reference frame of the flame front. All functions involved are then time-independent, and the argument $t$ can be omitted. In the absence of external gravitational field, the on-shell fresh gas velocity satisfies the following complex integro-differential equation \cite{kazakov1,kazakov2}
\begin{eqnarray}\label{master1}&&
2\omega_-' + \left(1 +
i\hat{\EuScript{H}\,}\right)\left\{[\omega]' -
\frac{Nv^n_+\sigma_+\omega_+}{v^2_+} \right\} = 0\,,
\end{eqnarray}
\noindent where $\omega = u + iw$ is the complex velocity, $[\omega] = \omega_+ - \omega_-$ its jump across the front; the subscript $- (+)$ denotes restriction to the front of a function defined upstream (downstream) of the front, {\it e.g.,} $w_-(x) = w(x,f(x)-0)$; the prime denotes $x$-differentiation, $N = \sqrt{1 + f'^2};$ $v^n = (\bm{v}\cdot\bm{n})$ is the normal gas velocity (the unit vector $\bm{n}$ normal to the front points to the burnt gas), $\sigma=\partial u/\partial x - \partial w/\partial y$ is the vorticity; finally, the operator $\hat{\EuScript{H}\,}$ is defined on $2b$-periodic functions by
\begin{eqnarray}\label{hcurvedf}
\left(\hat{\EuScript{H}\,}a\right)(x) = \frac{1 + i
f'(x)}{2b}~\fint\limits_{-b}^{+b}
d\eta~a(\eta)\cot\left\{\frac{\pi}{2b}(\eta - x +
i[f(\eta) - f(x)])\right\}\,,
\end{eqnarray}
\noindent the slash denoting the principal value of integral.

\begin{figure}
\centering
\includegraphics[width=0.5\textwidth]{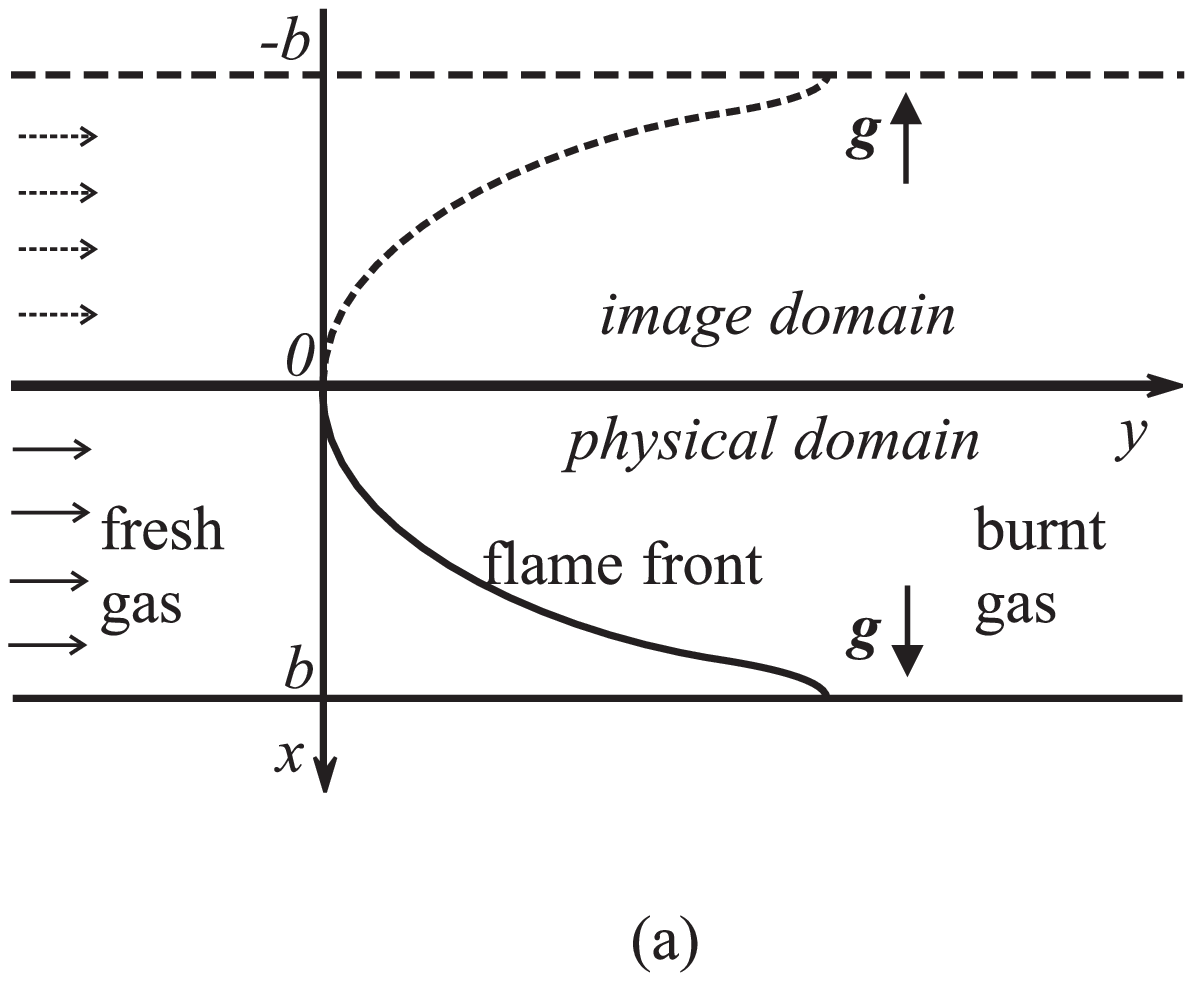}\vspace{2cm}
\includegraphics[width=0.5\textwidth]{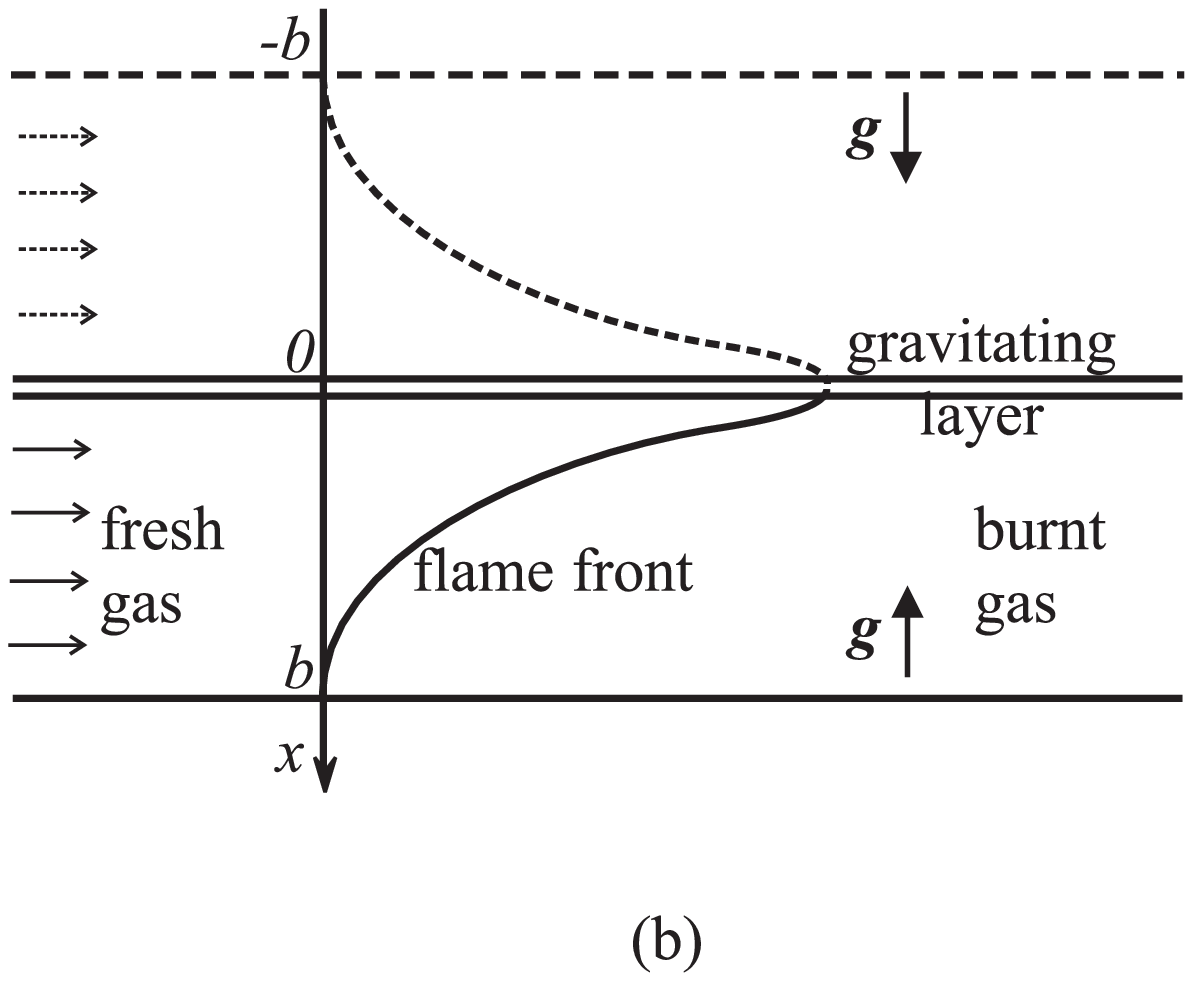}
\caption{}\label{fig1}
\end{figure}

Derivation of Eq.~(\ref{master1}) is based on a periodic continuation of the flame pattern along the $x$-axis, which proceeds in two steps: first, using the boundary conditions $w(0,y)=f'(0)=0$ (which express impermeability and ideality of the upper wall), the given flame pattern is carried from the physical domain $x \in [0,b]$ over to $x \in [-b,0]$ by reflection with respect to the $y$-axis. The resulting symmetric pattern satisfying
\begin{eqnarray}\label{reflect}
f(x) = f(-x),\quad w(x,y) = - w(-x,y), \quad u(x,y) = u(-x,y)
\end{eqnarray}
\noindent
is then periodically continued along the whole $x$-axis (using the same boundary conditions at the other wall). This trick is very convenient as it allows one to avoid explicit account of the boundary conditions, but it is physically inadequate in the presence of gravitational field perpendicular to the walls. This is because reflection performed at the first step leads to a discontinuity in the gravitational field -- its strength, $\bm{g},$ being directed downward in the physical domain, turns out to be directed upward in the image domain, Fig.~\ref{fig1}(a). In itself, this discontinuity is not a problem, because whether or not given configuration is physically relevant depends on its principal feasibility, and in the present case it is not difficult to find a setting realizing  physical conditions of the symmetric configuration. Namely, occurrence of a discontinuity in $\bm{g}$ suggests that the terrestrial gravitational field is to be replaced in the symmetric case by the field of a thin dense layer located near the plane $x=0.$ Taking its surface density equal to $g/2\pi G$ ($g\equiv |\bm{g}|,$ $G$ is the Newton gravitational constant) then yields gravitational field of the same strength as terrestrial, but opposite to it in the physical domain, so that the proper symmetric pattern is as shown in Fig.~\ref{fig1}(b). In this setting, the true flame configuration is obtained by a vertical translation from the image domain, $(-b,0) \ni x \to x + b,$ or by flipping the physical domain, $(0,b) \ni x \to b - x.$ Moreover, one can avoid this technical complication by introducing an electrically charged layer instead of the gravitating one, and assuming that the fresh and burnt gases are uniformly charged like the layer, with vanishingly small (to neglect electric interaction of different gas elements) charge densities standing in the same ratio as the mass densities, {\it viz.,} $\theta.$ But in any case, the symmetric configuration cannot be periodically continued in the $x$-direction, because of interference of the fields produced by the layers: Since the field of each additional layer is homogeneous, it cannot be neglected in the physical domain, which makes the continuation meaningless.

Fortunately, the on-shell description can be reformulated in a way that permits application to symmetric flame configurations independently of whether they admit periodic continuation. This reformulation was developed in \cite{jerk4} for flames propagating in symmetric channels with arbitrarily curved walls. It uses the method of Green functions to explicitly take into account boundary conditions at the walls, which leads to the following modification of the master equation
\begin{eqnarray}\label{master2}&&
2\omega_-' + \left(1 +
i\hat{\EuScript{H}\,}\right)\left\{[\omega]' -
\frac{M\omega_+}{v^2_+} +\frac{1+if'}{2b}\int_{-b}^{+b}d\eta\frac{M(\eta)\omega_+(\eta)}{v^2_+(\eta)}\right\} = 0\,,
\end{eqnarray}
\noindent where $M \equiv Nv^n_+\sigma_+.$ This equation differs from Eq.~(\ref{master1}) by the last term in the braces, which is proportional to the spatial average of the second term across the channel. As was noticed in \cite{jerk4}, the structure of the new term suggests that it describes an effective gravitational field along the channel, which appears in the system as a result of the curved flame evolution. For steady flame propagation in straight channels, such effective gravity presumably should vanish, for the opposite would mean that the flame is self-accelerating. But it turns out difficult to prove this directly, {\it i.e.,} to derive this vanishing from Eq.~(\ref{master1}). As to the real part of $\int^{+b}_{-b} d\eta M\omega_+/v^2_+,$ it does vanish as the average of an odd periodic function. However, its imaginary part is even, and there is no apparent reason why it should be zero. Using the small $(\theta-1)$-expansion, it can be shown that this quantity is indeed zero at least to the fourth post-Sivashinsky approximation (which corresponds to retaining terms of the seventh order in $(\theta-1)$ in the master equation) \cite{kazakov5}, but the general proof is still lacking. We return to this issue in Secs.~\ref{ident}, \ref{conclusions}.

Equation (\ref{master2}) will be used below in the study of uniform flame movement.

\subsection{Large-slope expansion}\label{largeslope}

It was already mentioned  that the speed of uniform movement in wide horizontal tubes largely exceeds the normal flame speed. Experiments indicate that flames propagating in tubes of diameter $d\lesssim 10$\,cm have fairly smooth elongated fronts, in which case flame speed is directly proportional to the front area. In wider tubes, flames develop complex unsteady small-scale structures observed as a widening of the front. The envelope surface retains highly elongated shape, but now part of the velocity increase is due to the small-scale wrinkling, which according to \cite{coward1932II} adds $40-50\%$ to the speed of uniform movement in comparison with the smooth patterns.

It follows that in the corresponding 2D case of flame propagation in channels of moderate\footnote{From the standpoint of dimensionless criterion $gb/U^2_f,$ all channels with $b > U^2_f/g$ ought to be qualified as wide. In the present investigation, however, it is more convenient to associate this term with the onset of instability of smooth flame configuration.} width ($b\lesssim$10\,cm), the front slope is large, $|f'|\gg 1,$ everywhere across the channel except narrow regions near the walls, where the front slope drops down to zero. The width of these regions is of order of the cutoff wavelength, $\lambda_c,$ related to the finite front-thickness effects which determine flame structure near the front end-points. $\lambda_c$ is generally larger for slower flames, but in no case does it exceed a few millimeters, whereas $b\geqslant$2.5\,cm in the experiments considered below. Thus, flame propagation in channels of moderate width can be adequately studied within the large-slope approximation. In wider channels ($b\gtrsim$10\,cm), condition $|f'|\gg 1$ is violated by the small-scale wrinkling, remaining valid only on average. As will be proved in Sec.~\ref{widechan}, the results obtained for smooth fronts hold true for wrinkled flames described in terms of averaged flow variables, which makes the large-slope approximation universally applicable to flame propagation in channels of any width.

A fact of fundamental importance for implementation of the large-slope approximation is the locality of steady flame patterns at large flow velocities, first discovered  in the study of anchored flames \cite{kazakov3}. This locality means that in regions where $|f'|\gg 1,$ front curvature at any point is uniquely determined by the front position and its slope at the same point. A similar statement is true of the on-shell gas velocity. In other words, these functions satisfy ordinary differential equations.\footnote{In this respect, the term ``locality'' should not be understood as independence of distant flame elements of each other. On the contrary, the asserted locality implies that the whole flame pattern can be restored (by solving an ordinary differential equation) from the knowledge of $f$ and $f'$ at any point. In this way, a nonlocal impact of flame anchors was described in \cite{kazakov3}.} The underlying mathematical reason  is that the operator $\hat{\EuScript{H}\,},$ which mainly determines the nonlocal structure of the master equation, becomes localizable when the front slope is large. Namely, its large-slope expansion reads, in the natural units (we consider henceforth the symmetric pattern shown in Fig.~\ref{fig1}(a))
\begin{eqnarray}\label{hcurvedf1}
\left(\hat{\EuScript{H}\,}a\right)(x) = (f'(x) -
i)~\int\limits_{0}^{1} d\eta~\frac{a(\eta) +
a(-\eta)}{2}\chi(\eta - |x|)+ia(-x)(2|x| - 1) +
O\left(\frac{1}{f'}\right)\,,
\end{eqnarray}
\noindent where $\chi(x)$ is the sign function,
$$\chi(x) = \left\{
\begin{array}{cc}
+1,& x>0\,,\\
-1,&  x<0\,.
\end{array}
\right.
$$ It is because of the step-like asymptotic form of the integral kernel of $\hat{\EuScript{H}\,}$ that the master equation reduces eventually to an ordinary differential equation. The other possible source of non-locality is the last term in Eq.~(\ref{master2}), which depends on the values of $f'(x),u_-(x),w_-(x)$ at every point of the front, including the regions where the large-slope approximation is not valid. A remarkable fact to be proved below is that this term does not prevent localization, falling off eventually from the master equation.

The following version of Eq.~(\ref{hcurvedf1}), expressing the action of $\hat{\EuScript{H}\,}$ on a derivative, is also useful:
\begin{eqnarray}\label{hcurvedf2}
\left(\hat{\EuScript{H}\,}a'\right)(x) = (f'(x) -
i)\left\{a(-|x|) - a(|x|)\right\} + ia'(-x)(2|x| - 1) +
O\left(\frac{1}{f'}\right),
\end{eqnarray}
\noindent where the prime denotes derivative of the function
with respect to its argument, $a'(y) = da(y)/dy.$ An important result proved in \cite{kazakov3} is that this formula holds true for functions allowed to have a discontinuity at the origin.

\subsubsection{Evolution equation and dispersion relation. Boundary conditions}\label{boundarycond}

The local rate of consumption of fresh gas is determined by the so-called evolution equation which expresses the normal fresh gas velocity relative to the front as a local functional of the front slope and the on-shell gas velocity. In view of this locality, the evolution equation for zero-thickness flames is that for planar flames, {\it viz.,} $v^n_- - \dot{f}/N= U_f$ (dot denotes time differentiation), whereas account of the transport processes inside the front adds to the right-hand side terms proportional to the front thickness. The relative value of these terms is of order $\lambda_c/\lambda,$ where $\lambda$ is the wavelength of the front wrinkling. Therefore, they do not affect flame propagation in channels of moderate width. On the other hand, it was already mentioned that the results obtained for smooth patterns apply also to wrinkled flames upon averaging over the small-scale structure. Hence, it will be sufficient to use the master equation along with the evolution equation for zero-thickness steady flames, which in terms of the Cartesian velocity components reads, in the natural units,
\begin{eqnarray}\label{evolutiongen}
u_- - f'w_- = N\,.
\end{eqnarray}
\noindent In the leading order of the large-slope expansion, it can be written as
\begin{eqnarray}\label{evolutiongen1}
u_- = f'(w_- + 1).
\end{eqnarray}
\noindent This is for $f'>0,$ {\it i.e.,} $x>0$ in the configuration of Fig.~\ref{fig1}(a); hereon, whenever a local relation appears, it applies to the physical domain.

Let $U$ denote the steady flame speed with respect to the channel. In the rest frame of the flame front, $U$ is the velocity of incoming fresh gas. Hence, the longitudinal component of the on-shell gas velocity, $u_-,$ is $O(U).$ Next, the total mass conservation implies that
\begin{eqnarray}\label{totalmass}
U = \int\limits_{0}^{1}d\eta \sqrt{1 + f'^2}\, v^n_- \approx \int\limits_{0}^{1}d\eta f' = f(1) - f(0)\,.
\end{eqnarray}
\noindent
Therefore, of the same order is the (smooth) front slope, $f' = O(U).$ Equation (\ref{evolutiongen1}) then implies that the other (transversal) velocity component $w_- = O(1).$ An equation relating the three functions can be obtained from the following dispersion relation \cite{kazakov2}
\begin{eqnarray}\label{chup}
\left(1 - i \hat{\EuScript{H}\,}\right)\left(\omega_-\right)' &=& 0\,,
\end{eqnarray}
\noindent which is a consequence of potentiality, incompressibility and boundedness of the flow upstream. This equation is actually contained in Eq.~(\ref{master2}), which can be seen by applying the operator $(1 - i \hat{\EuScript{H}\,})$ to its left-hand side and using the identity $\hat{\EuScript{H}\,}^2 = - 1$ \cite{kazakov2}. But within the large-slope expansion employing expression (\ref{hcurvedf1}) or (\ref{hcurvedf2}), Eq.~(\ref{chup}) must be used separately. The point is that the expansions of $\EuScript{H}$-operator, given in the preceding section, allow only extraction of the leading terms in Eqs.~(\ref{master2}), (\ref{chup}). This is directly related to the fact that the exact identity $\hat{\EuScript{H}\,}^2 = - 1$ cannot be verified using these expansions, for the composition of the leading (integral) term in Eq.~(\ref{hcurvedf1}) with the undetermined remainder is $O(1).$

To extract the leading contribution from Eq.~(\ref{chup}), it is to be noted that  the action of $\hat{\EuScript{H}\,}$ on a flow variable essentially depends on its parity properties. Namely, Eq.~(\ref{hcurvedf2}) shows that $\hat{\EuScript{H}\,}a' = O(1)$ if the function $a(x)$ is even (assuming that $a = O(1)$), and $\hat{\EuScript{H}\,}a' = O(U),$ if it is odd. Recalling Eq.~(\ref{reflect}) and the above estimates for $u_-,w_-,$ one readily sees that the leading contribution to the left-hand side of Eq.~(\ref{chup}) is real. We thus obtain the following relation $(x>0)$
\begin{eqnarray}\label{3rel}
f'(x)w_-(x) = (1 - x)u_-'(x)\,.
\end{eqnarray}
\noindent Its validity is conditioned by that of Eq.~(\ref{hcurvedf2}), which requires the function $f(x)$ be monotonic in the physical domain, and have large slope. As we know, the latter does not hold near channel walls. This is not actually a problem near the wall $x=1,$ for $f' = 0$ is not a fundamental requirement in the case of zero-thickness flames. It appears only upon account of the finite front-thickness effects, which raises differential order of the equation for the front position, calling thereby for an additional boundary condition. Note that Eq.~(\ref{3rel}) is satisfied at the point $x=1$ even though $f'(1)\ne 0,$ by virtue of the boundary condition $w_-(1)=0.$ It is therefore legitimate to discard the narrow region where $f'$ drops down to zero, and to use Eq.~(\ref{3rel}) also near the wall $x=1,$ which will be assumed henceforth. Things are different, however, near the wall $x=0.$ Although conditions $f'(0) = 0$ and $f'(1) = 0$ have the same status from the point of view of their genesis just described, differentiability of the flow variables everywhere within the extended channel $x\in (-1,+1)$ is essentially used in the derivation of Eq.~(\ref{chup}) and of the master equation. In fact, it is existence of the short-wavelength cutoff, $\lambda_c,$ that guarantees analyticity of the complex velocity, used to derive the dispersion relation. This analyticity would be lost if the front had an edge, which necessarily occurs at $x=0$ if $f'(0)\ne 0.$

Thus, self-consistency of the approach requires vanishing of odd functions at the origin: $f'(0) = w_-(0)=u'_-(0)=0.$ But the narrow region near $x = 0,$ where these functions rapidly turn into zero, is still of no interest. To avoid the necessity of explicit consideration of this region, we shall proceed as follows. We note that in view of locality of the steady flame structure, announced in the preceding section, initial conditions for the ordinary differential equations satisfied by $f'(x),$ $u_-(x),$ $w_-(x)$ can be specified at any internal point of the physical domain where the front slope is large, say, $x=0.5.$ The values of the slope and the on-shell velocity at this point determine the whole flame pattern.\footnote{It should be stressed that in general, these values are functionals of the entire flame structure, and in particular, of the flame properties near the walls. But it will be shown later on that within the approximations used, they are actually independent of the near-the-wall flame behavior (see Sec.~\ref{reduction}).} Let us now think of the functions $f'(x),$ $u_-(x),$ $w_-(x)$ as satisfying Eqs.~(\ref{evolutiongen1}), (\ref{3rel}) for all $x\in (0,1),$ including the regions near the walls. In other words, we use equations obtained within the large-slope expansion to continue their solution to the whole physical domain. It will be seen in the next section that the continuation is unique by virtue of the uniqueness theorem for ordinary differential equation resolved with respect to the higher derivative. Thus, the original configuration and the continued one coincide everywhere except in a vicinity of $x=0,$ where the true functions $f'(x)$ and $w_-(x)$ rapidly turn into zero, but the continued ones do not. To avoid proliferation of symbols, the current notation will be reserved for the functions continued in the way just described (in what follows, the original functions appear again only once -- in formulas (\ref{master3}) to (\ref{master4}), so no confusion should arise).

As said, any point with a large front slope can be used for specifying initial conditions, but the end-points $x=0$ and $x=1$ are, of course, most suitable, and they are now admissible for this purpose, as the continued flame pattern is local in the whole physical domain. Let $w_0$ and $f'_0$ denote the limiting values for $x\to 0^+$ of the functions $w_-(x)$ and $f'(x),$ respectively:\footnote{The same notation $w_0$ appears also in the study of anchored flames \cite{kazakov3}, but with a different meaning: $w_0$ denotes presently the boundary value of the continued velocity, whereas in the case of a flame anchored by a rod of vanishingly small radius it represents the true gas velocity induced by the rod.}
$$w_0 \equiv w_-(0^+), \quad f'_0 \equiv f'(0^+).$$ These parameters, introduced for later convenience, must be expressed eventually in terms of the physical parameters of the theory -- $\theta,U,g.$ But one relation can be inferred already now from the evolution equation. To this end, we have to identify boundary conditions for the functions $f(x),u_-(x).$ First, it is a mere convention to set
\begin{eqnarray}\label{bcondition1}
f(0) = 0,
\end{eqnarray}
\noindent but in conjunction with Eq.~(\ref{totalmass}) it implies that
\begin{eqnarray}\label{bcondition2}
f(1) = U.
\end{eqnarray}
\noindent
The last condition follows from potentiality of the upstream flow: By virtue of the Thomson theorem, one has $\sigma = 0$ there, which implies that $$\partial u/\partial x = \partial w/\partial y = O(1).$$ Hence, to the leading order $u(x,y) = U$ for $y\leqslant 0,$ in particular, $u(0,0)=U,$ or
\begin{eqnarray}\label{bcondition3}
u_-(0) = U.
\end{eqnarray}
\noindent Now, setting $x=0$ in Eq.~(\ref{evolutiongen1}), the latter condition gives
\begin{eqnarray}\label{bcondition0}
f'_0 = \frac{U}{w_0 + 1}\,.
\end{eqnarray}
\noindent
We mention finally that there is a simple algebraic relation between the functions $f(x)$ and $u_-(x)$ \cite{kazakov3}. Namely, eliminating $w_-$ from Eqs.~(\ref{evolutiongen1}), (\ref{3rel}) gives $u_- = (1-x)u'_- + f',$ or $[(1-x)u_-]' + f' = 0\,.$ Integration of this equation under condition (\ref{bcondition2}) yields
\begin{eqnarray}\label{main3}
f = U - (1 - x)u_-\,.
\end{eqnarray}
\noindent

\subsubsection{Master equation in the leading order. Reduction to ordinary differential equation}\label{reduction}

Let us turn to the master equation. The velocity jumps across the front and the on-shell vorticity, entering this equation, read in the natural units
\begin{eqnarray}\label{jumps}
v^n_+ &=& \theta\,, \quad [u] = \frac{\theta - 1}{N}\ ,
\quad [w] = - f'\frac{\theta - 1}{N}\,, \\
\sigma_+ &=& - \frac{\theta - 1}{2\theta N}(u^2_- + w^2_-)' + \frac{\theta - 1}{\theta N}g\chi(x)\,.\label{vorticity}
\end{eqnarray}
\noindent The last term in Eq.~(\ref{vorticity}) describes the baroclinic effect \cite{hayes}. As in the preceding section, the order-of-magnitude estimates for the flow variables together with their parity properties imply that the leading contribution to the left-hand side of Eq.~(\ref{master2}) is real. Extracting it with the help of Eq.~(\ref{hcurvedf1}), we find
\begin{eqnarray}\label{master3}
2u'_-(x) (1 + \alpha |x|)- \frac{2\alpha g x}{u_-(x)} - \gamma f'(x) +
f'(x)\int\limits_{0}^{1}d\eta \left[\frac{Mw_+}{v^2_+} +\alpha\left(\frac{f'}{N}\right)'\right]\!\!(\eta)\,\chi(\eta - |x|) = 0, \nonumber\\
\end{eqnarray}
\noindent
where $\alpha \equiv \theta - 1,$ and $\gamma$ denotes the average value of $Mw_+/v^2_+$ across the channel,
\begin{eqnarray}\label{gamma}
\gamma = \int\limits_{0}^{1}d\eta \frac{M(\eta)w_+(\eta)}{v^2_+(\eta)}\,.
\end{eqnarray}
\noindent Just like Eq.~(\ref{3rel}), Eq.~(\ref{master3}) is valid only for $\{x: |f'(x)|\gg 1\}.$ By this reason, the large-slope expansion is not applicable to the integrand, as it involves the region near $\eta = 0.$ First of all, let us evaluate the integral term with $(f'/N)'.$ In the most part of the channel the slope is large, so that $f'/N\approx 1$ in the physical domain, and $f'/N\approx - 1$ in the image domain. Therefore, the function $(f'/N)'(x)$ is localized near $x=0$ (as discussed in Sec.~\ref{boundarycond}, the slope can be consistently assumed large at the wall $x=1,$ despite the actual flattening caused by the finite front-thickness effects). Using this, and taking into account that $(f'/N)'(x)$ is an even function, one can write
$$\int\limits_{0}^{1}d\eta \left(\frac{f'}{N}\right)'\!\!(\eta)\,\chi(\eta - |x|) = \frac{1}{2}\int\limits_{-1}^{1}d\eta \left(\frac{f'}{N}\right)'\!\!(\eta)\,\chi(- |x|) = - \frac{1}{2}\left.\frac{f'}{N}\right|_{-1}^{+1} = -1.$$
Equation (\ref{master3}) thus takes the form
\begin{eqnarray}\label{master4}
2u'_-(x) (1 + \alpha |x|)- \frac{2\alpha g x}{u_-(x)} - \alpha f'(x) - \gamma f'(x) + f'(x)\int\limits_{0}^{1}d\eta \frac{Mw_+}{v^2_+}(\eta)\,\chi(\eta - |x|) = 0\,.
\end{eqnarray}
\noindent
Dividing it by $f',$ and differentiating with respect to $x$ leads to an
ordinary differential equation ($x>0$)
\begin{eqnarray}\label{master5}
\frac{d}{dx}\left[\frac{u'_-}{f'} (1 + \alpha x) - \frac{\alpha g x}{f'u_-}\right] + \alpha\left(\frac{u'_-}{u_-} - \frac{g}{u^2_-}\right)(w_- - \alpha) = 0\,.
\end{eqnarray}
\noindent This proves the asserted locality of the steady flame structure, since together with Eqs.~(\ref{evolutiongen1}) and (\ref{3rel}), Eq.~(\ref{master5}) constitutes a closed system of ordinary differential equations for the functions $f(x),$ $u_-(x),$ $w_-(x).$

As derived, Eqs.~(\ref{master3})--(\ref{master5}) apply for $\{x: |f'(x)|\gg 1\},$ and involve functions which describe the true flow in the whole channel (including the region near $x=0$). We now extend Eq.~(\ref{master5}) to all $x\in (0,1).$ As explained in the preceding section, Eqs.~(\ref{evolutiongen1}), (\ref{3rel}), and (\ref{master5}), considered on the interval $x\in (0,1),$ define continuation of the flow variables to the whole physical domain. It remains only to infer an additional initial condition for the continued functions $f'(x),u_-(x),$ to account for the rise of differential order in the transition from Eq.~(\ref{master4}) to Eq.~(\ref{master5}). To this end, we note that Eq.~(\ref{master4}) is still satisfied by the continued functions, if it is understood that the integral term and the constant $\gamma$ are still evaluated on the true functions. Letting $x \to 0$ then gives $2u'_-(0^+) - \alpha f'(0^+) = 0$ or, using Eq.~(\ref{3rel}),
\begin{eqnarray}\label{bcondition4}
w_0 = \frac{\alpha}{2}\,.
\end{eqnarray}
\noindent This result deserves some commentary. The dimensionless quantity $w_0$ might in principle depend on the three independent dimensionless parameters of the theory -- $U,\theta,$ and $g$ ($= gb/U^2_f,$ in the ordinary units). It is independent of $U$ merely because the large-slope asymptotic expansion we use is, in fact, an expansion in powers of $U,$ whereas $w=O(1).$ Next, independence of $g$ is accidental: There is no reason to expect that it will hold in the subleading order [independence of $g$ would be exact if $w_0$ were determined entirely by the flow properties in the narrow region near $x=0,$ since gravity is negligible on small scales (as is evident from the fact that gravitational effects are described by terms of the lowest differential order, Cf. {\it e.g.,} Eq.~(\ref{vorticity})). However, $w_0$ is the boundary value of velocity continued from within the physical domain using the differential equation which involves $g$]. What is remarkable is that $w_0$ turned out to be independent of $\gamma,$ and hence, of the true flow specifics near the walls. This independence makes the present consideration self-contained, in that it does not require the knowledge of $\gamma$ to describe flame structure far from the walls. This conclusion is not actually quite unexpected, as we deal here with flames evolving spontaneously, in contrast to flames with a fixed ignition point, where large gradients of the gas velocity near flame holder do contribute to $w_0,$ affecting thereby the whole flame pattern. Regarding the constant $\gamma$ itself, it should be stressed that its exact value cannot be found using the continued functions, precisely because they fail to describe the true flow near $x=0.$ Nevertheless, approximate values can be instructive, Cf. Secs.~\ref{ident}, \ref{conclusions}.

Finally, combining Eqs.~(\ref{evolutiongen1}), (\ref{3rel}) one finds
\begin{eqnarray}\label{auxeq}
f' = u_- - (1-x)u'_-\,, \quad w_- = \frac{(1-x)u'_-}{u_- - (1-x)u'_-}\,.
\end{eqnarray}
\noindent Substitution of these expressions into Eq.~(\ref{master5}) yields a second order differential equation for $u_-(x),$ which is linear with respect to $u''_-(x).$ Therefore, its solution is unique, if exists. The initial conditions for this equation follow from Eqs.~(\ref{bcondition3}), (\ref{bcondition0}), and (\ref{bcondition4})
\begin{eqnarray}\label{bcondition5}
u_-(0) = U\,, \quad u'_-(0) = \frac{\theta - 1}{\theta+1}\,U\,.
\end{eqnarray}
\noindent

At last, it is necessary to make the following important observation regarding general structure of equations derived in the leading order of the large-slope expansion. Dimensional analysis suggests that the flame speed can be written as
$$U = C\left(\theta,gb/U^2_f\right)\sqrt{gb}\,,$$ where $C$ is a function of the two dimensionless combinations of parameters characterizing the problem under consideration. Since the flame speed grows with $g$ (for it is gravity that forces flame to spread along the channel), taking the large-slope limit implies that $C(\theta,gb/U^2_f)$ is replaced by $C(\theta,\infty),$ {\it provided that the limit of $C$ exists.} That the limit should exist is suggested by the observation that $g$ disappears from Eqs.~(\ref{master5}), (\ref{auxeq}) on replacing
$$u_-\to \sqrt{g}\,u_-, \quad f\to \sqrt{g}f, \quad w_-\to w_-.$$ However, this suffices to prove existence of the limit only if the function $C\left(\theta,gb/U^2_f\right)$ is {\it single-valued.} The nontrivial fact established below is that this function actually is {\it not} single-valued, and this leads to interesting complications in the relationship between flame speed and tube diameter, to be discussed in Sec.~\ref{comparison}.

\subsubsection{Numerical solutions}\label{numeric}

\begin{figure}
\centering
\includegraphics[width=0.9\textwidth]{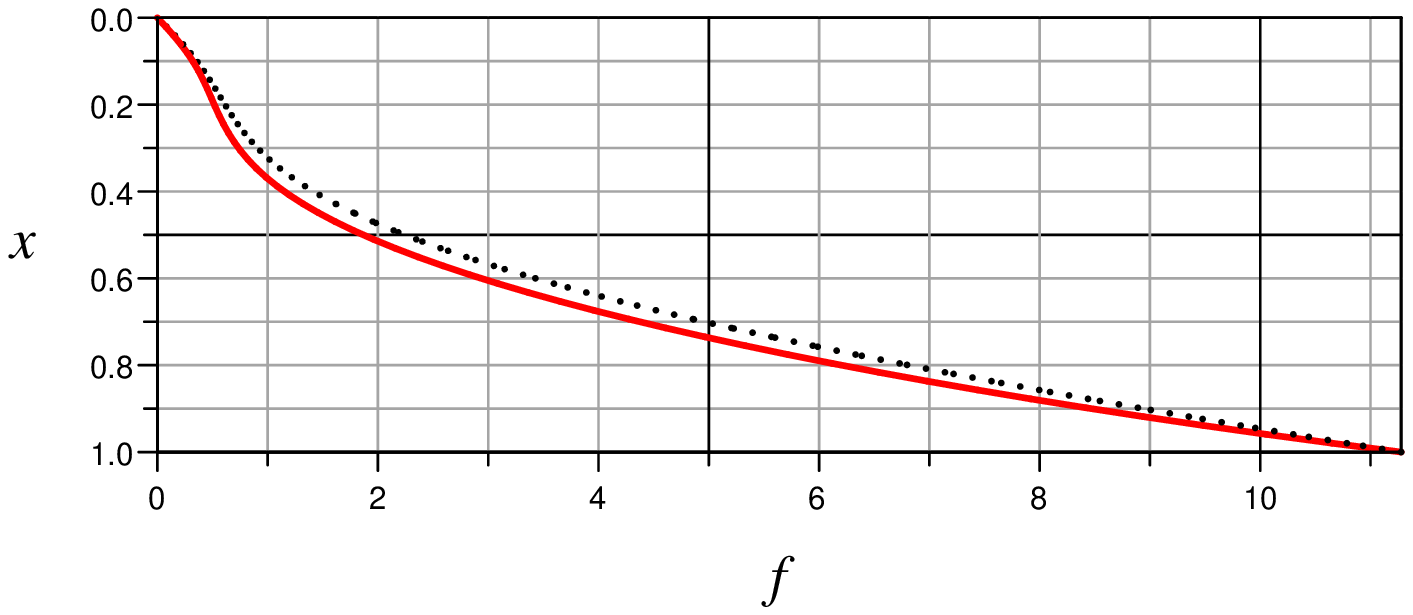}
\includegraphics[width=.4\textwidth]{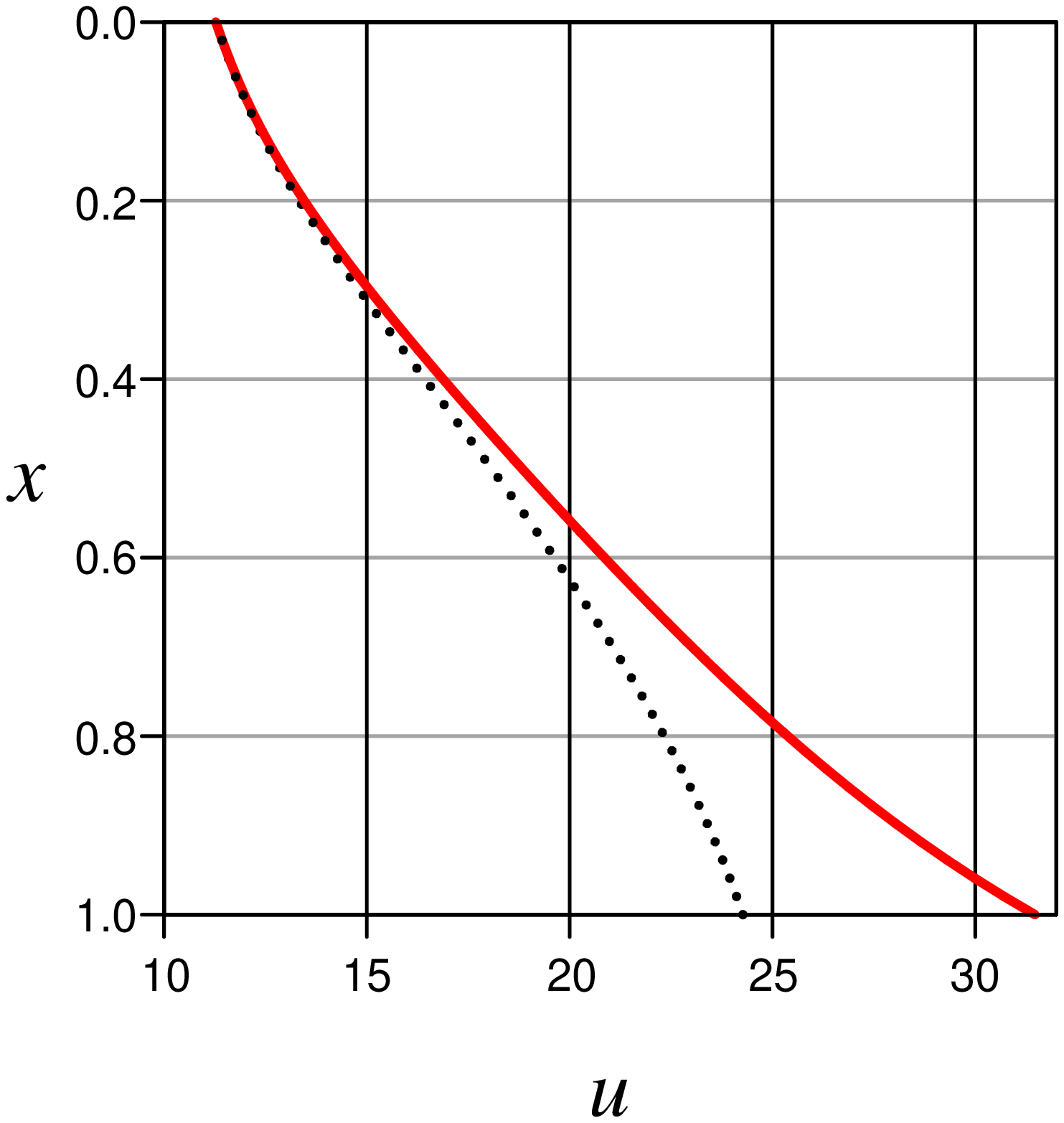}\hspace{1cm}
\includegraphics[width=.4\textwidth]{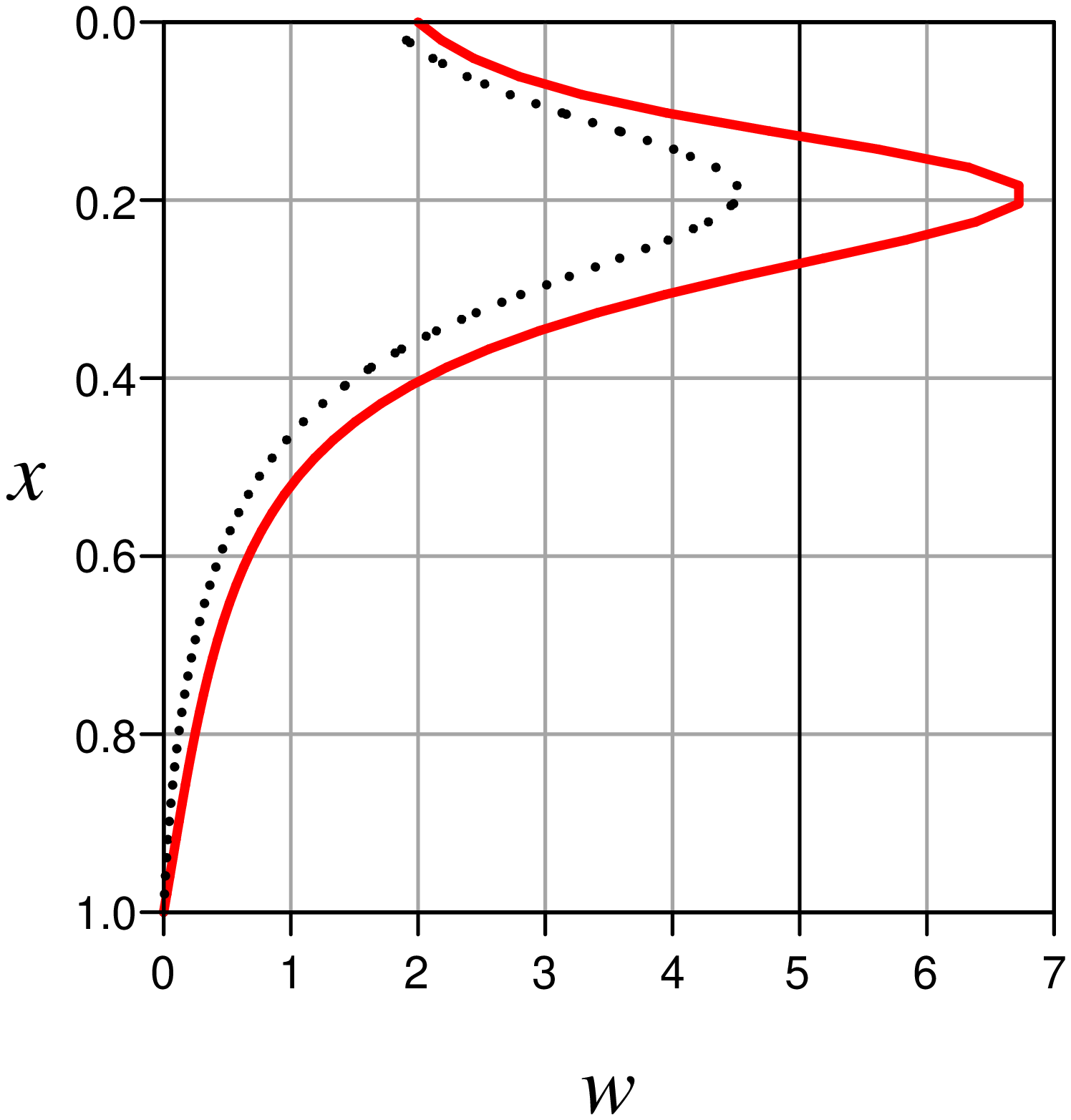}
\caption{}\label{fig2}
\end{figure}

\begin{figure}
\centering
\includegraphics[width=0.9\textwidth]{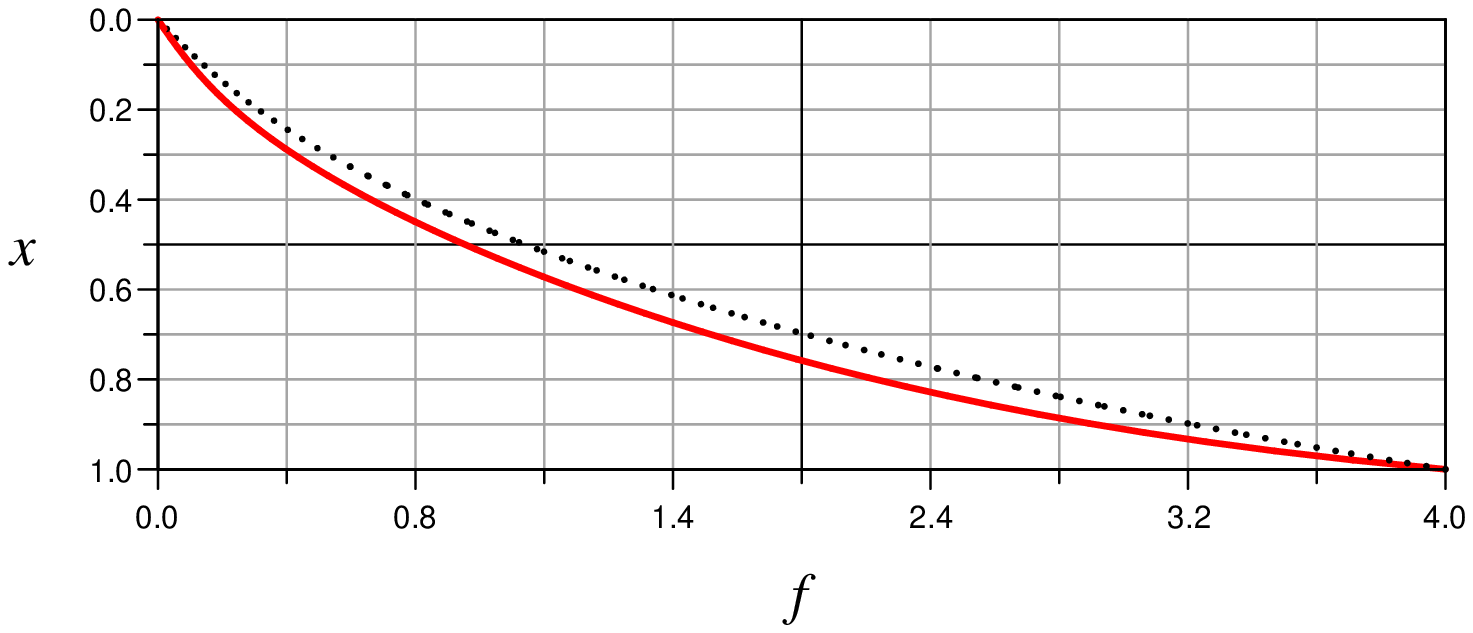}
\includegraphics[width=.4\textwidth]{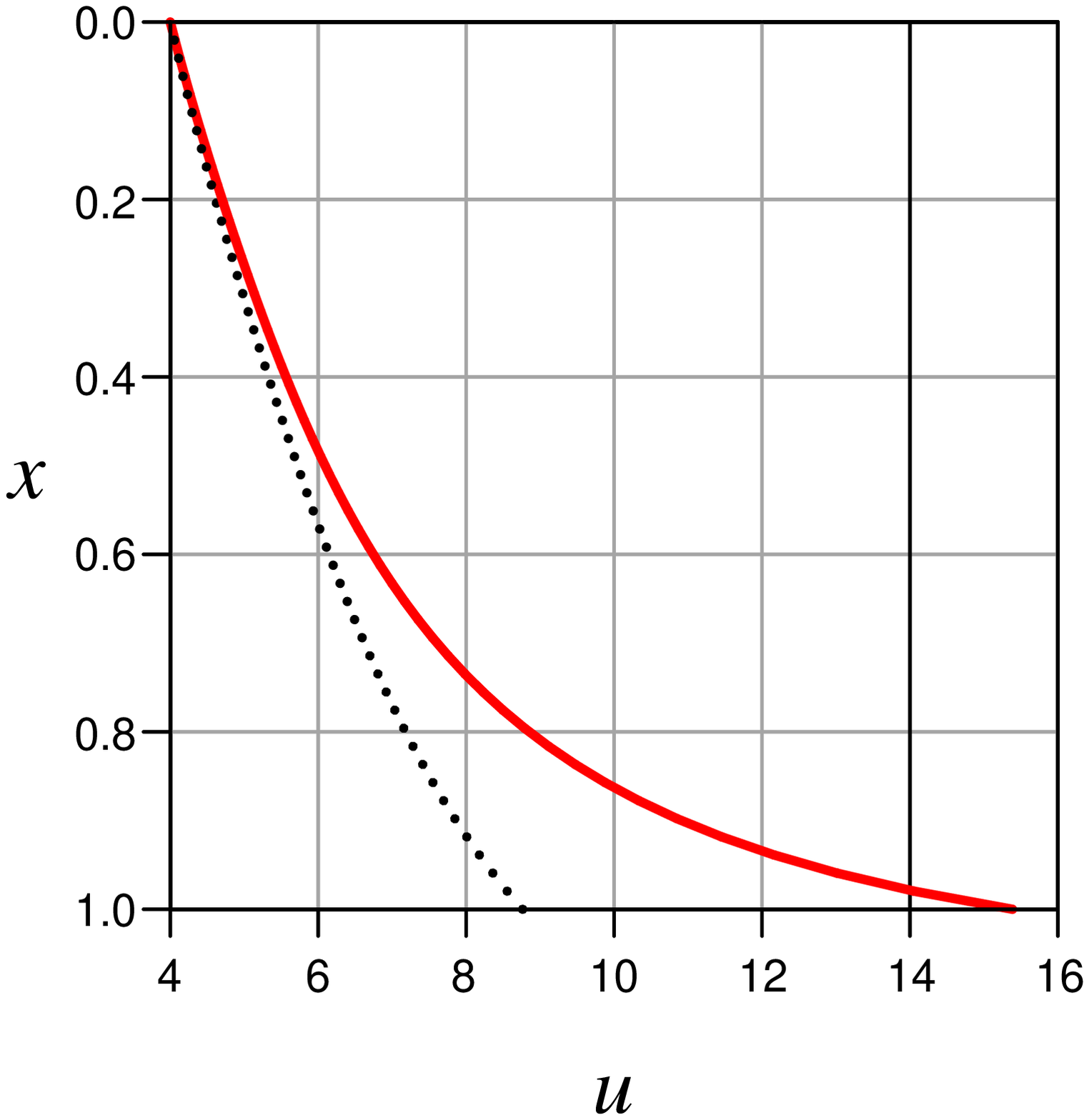}\hspace{1cm}
\includegraphics[width=.41\textwidth]{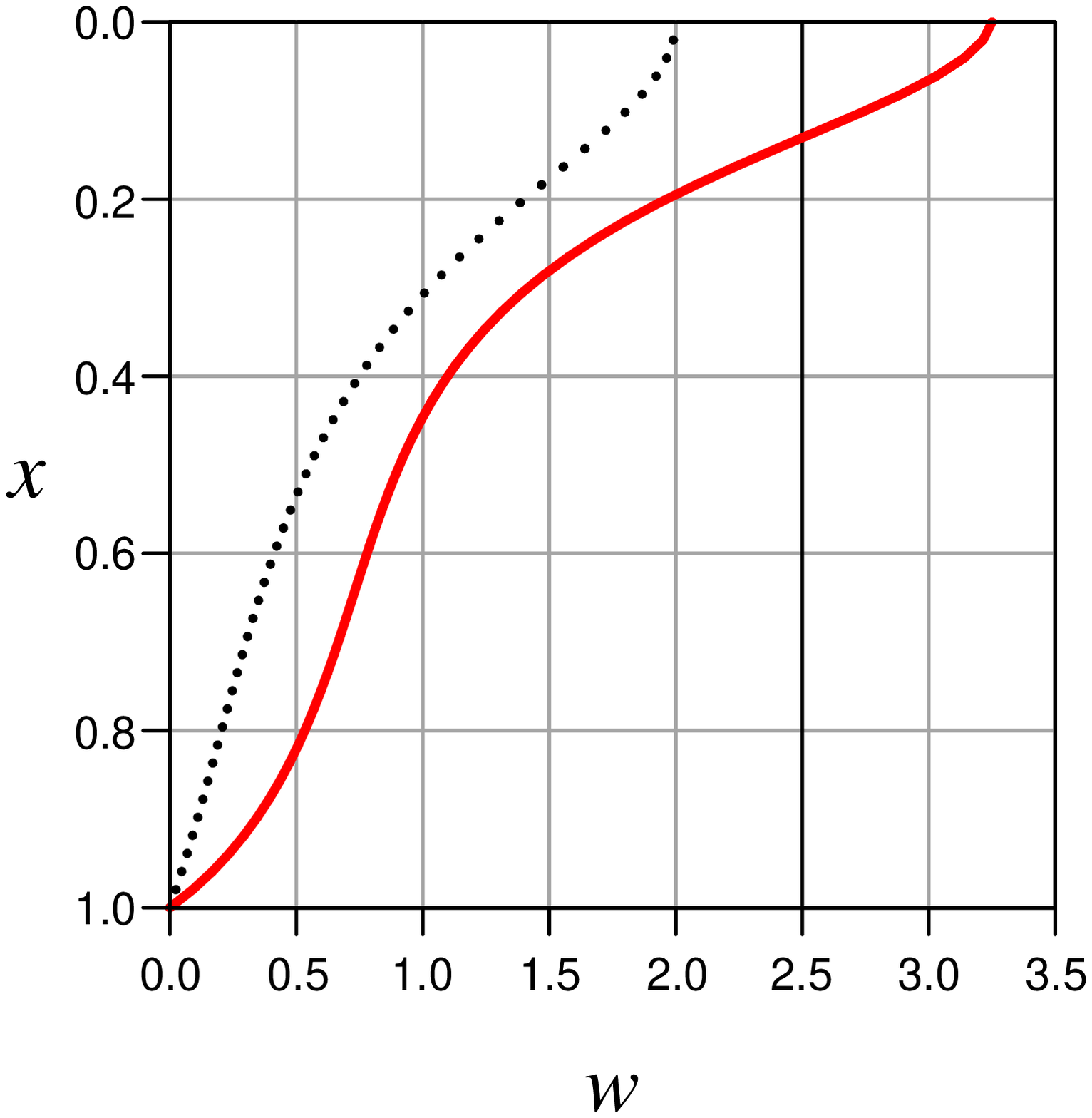}
\caption{}\label{fig3}
\end{figure}

It turns out that for each set of parameters $\theta,g,$ infinitely many solutions to the equation for $u_-(x)$ exist, which differ by the value of $U.$ Namely, there are two domains on the line $U>0,$ one of which is a finite interval $U\in (U_0,U_1),$ and the other is an infinite interval $U>U_2>U_1,$ where $U_{0,1,2}$ depend on $\theta,g.$ Henceforth, solutions having $U$ in these domains will be referred to as {\it Type I} and {\it Type II} solutions, respectively. Physical solutions of each type will be identified in Sec.~\ref{ident}, meanwhile some examples are given in order to characterize the two types of flame patterns. Figure \ref{fig2} shows front positions and on-shell fresh gas velocities for Type I flames with $\theta = 5$ and $\theta =4.5,$ propagating with a normal speed $U_f=10$\,cm/s in a channel of $b=40$\,cm. The dimensionless speed of uniform movement in both cases is $U=11.28.$ Figure \ref{fig3} shows similar data for Type II flames with $\theta =7.5$ and $\theta=5,$ both having $U_f = 40$\,cm/s, $U=3.9,$ and propagating in a channel of $b=30$\,cm.
It is seen that in all cases longitudinal velocity monotonically increases with $x,$ which implies that the gas pressure drops downstream, as it normally does in actual experiments. The key difference between Type I and Type II solutions is in the behavior of transversal velocity: it changes monotonically across the channel in the latter case, and has a pronounced maximum in the former. This maximum corresponds to the inflection point on the flame front, which is characteristic of Type I solutions.

\section{Flame propagation in wide channels}\label{widechan}

\begin{figure}
\centering
\includegraphics[width=0.5\textwidth]{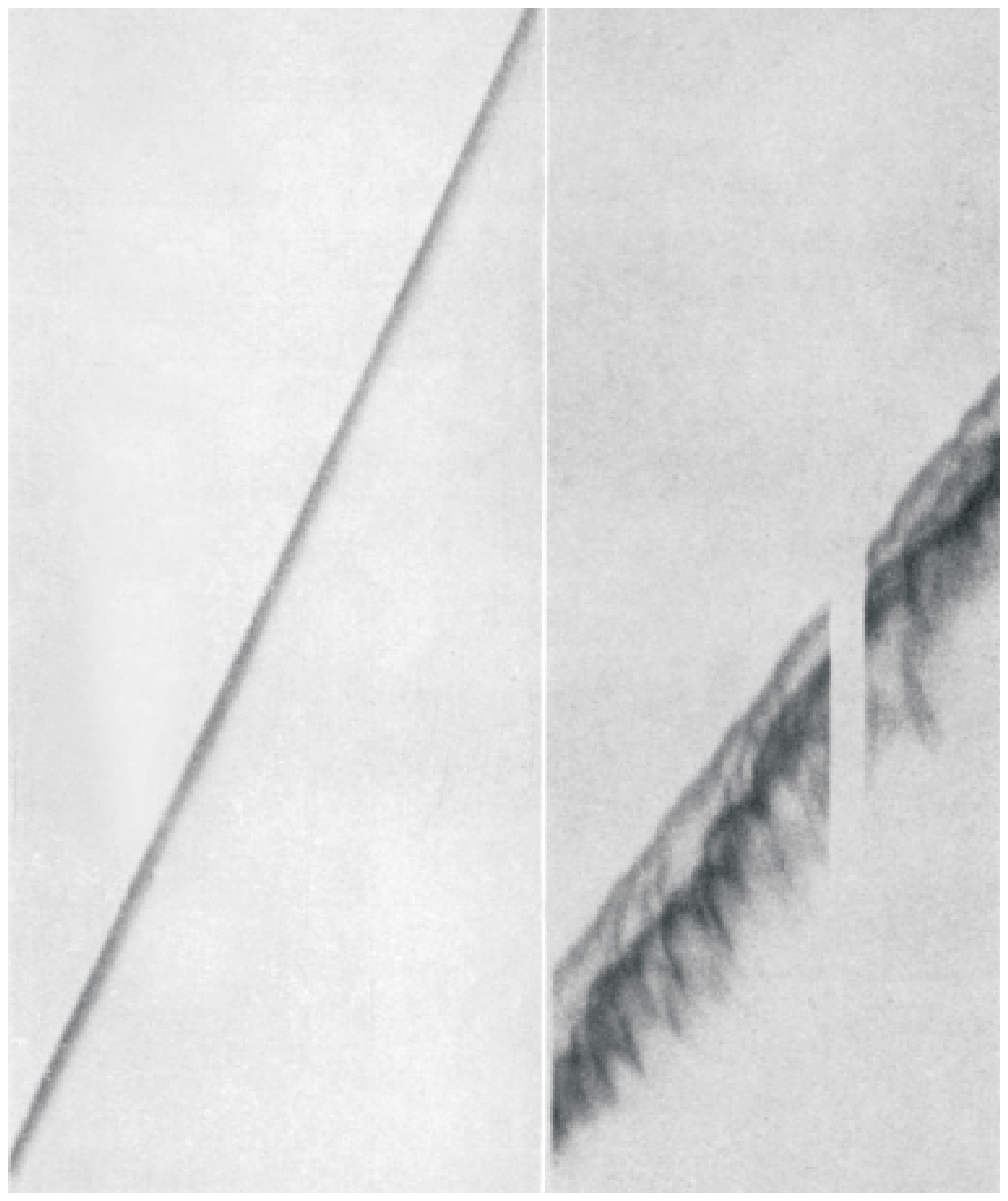}
\caption{}\label{fig4}
\end{figure}

Experiments \cite{mason1917,mason1923,coward1928,coward1932I,coward1932II} show that
the steady regime of flame propagation is not realized in sufficiently wide horizontal tubes. Instead, flames develop irregularly varying small-scale structures observed visually as a flame turbulence, or as a widening of the flame front, on photographs. For methane-air flames, for instance, the critical tube diameter $d\approx 10$\,cm. Mason and Wheeler \cite{mason1917} described this turbulence as ``a swirling motion in a direction nearly normal to the direction of translation of the flame-front,'' and attributed it to gas convection. Figure \ref{fig4} reproduces moving-film photographs of flames propagating in a $10\%$ methane-air mixture in tubes of $2.5$ and $20$\,cm diameter \cite{coward1932I}. It shows a straight-line front in the former case, and complex geometrical patterns present at every instant in the latter. It is very important, however, that the envelope surface (the average front position) retains its shape during the phase of uniform movement. Moreover, it is seen from the photograph that the speed of uniform movement remains grossly constant. These facts justify the very notion of uniform movement, and suggest that this phase of flame evolution may be described in terms of flow variables averaged somehow over the small-scale flame irregularities. An appropriate averaging procedure can indeed be introduced, which allows extension of the obtained results to wide channels at the cost of introducing only one additional phenomenological parameter -- the effective normal front speed describing amplification of the local consumption rate, caused by the small-scale front wrinkling. This will be done below following the lines of \cite{kazakov2003}, with several clarifications and a simplification of the averaging procedure which will be purely spatial (Sec.~\ref{aprocedure}). The main consideration consists of two parts dealing separately with the bulk equations (Sec.~\ref{eqdecoupling}), and the local propagation law (\ref{renorm}).

\subsection{Averaging procedure}\label{aprocedure}

Dynamics of flame disturbances characterized by the length scale $\lambda_c$ is governed by the finite front-thickness effects. These effects are the result of redistribution of heat and mass fluxes in the preheat zone of the flame front, caused by the front curvature and inhomogeneity of the tangential flow. The simple idea behind the averaging procedure to be introduced is that the strong local effects due to this redistribution must largely compensate each other when their influence on the flame dynamics at distances $\gg \lambda_c$ is considered. Using the condition $\lambda_c\ll b,$ let us choose a length $L$ satisfying
\begin{eqnarray}\label{scaleseparation}
\lambda_c \ll L \ll b\,.
\end{eqnarray}
\noindent Given a function $A(\bm{x},t),$ we define its spatial average
over the region $V = \{\bm{x}: \bm{x}\in (\tilde{\bm{x}},\tilde{\bm{x}} + \Delta\bm{x}),$
$\Delta x_i = L\}$ near a point $\tilde{\bm x}$ as
\begin{eqnarray}\label{average}
\langle A \rangle = \frac{1}{L^2} \int\limits_{V}d^2\bm{x}~A(\bm{x},t) \equiv A_0(\tilde{\bm{x}},t)\,.
\end{eqnarray}
\noindent By definition, $\langle A\rangle$ varies noticeably
over distances $\sim b.$ The function $A(\bm{x},t)$ thus turns out to be decomposed into two parts corresponding to the two scales
$b$ and $\lambda_c:$
\begin{eqnarray}\label{split}
A = A_0 + A_1, \qquad \langle A_1 \rangle = 0.
\end{eqnarray}
\noindent
Both the slowly varying ($A_0$) and rapidly varying ($A_1$) parts will be treated as power series in the small ratio $\lambda_c/b\equiv\varepsilon$
$$A_0 = A_0^{(0)} + \varepsilon A_0^{(1)} + \dots\,,
\quad A_1 = A_1^{(0)} + \varepsilon A_1^{(1)} + \dots\,,$$ where dots denote terms of higher order in
$\varepsilon.$ Despite formal resemblance of these expansions, their meaning is different. According to the above definition, dependence of $A_0$ on $\lambda_c$ is naturally expected to be regular enough to admit expansion in powers of $\lambda_c/b.$ This is because $\lambda_c$ is not the length scale of $A_0,$ and decreasing $\lambda_c$ only improves conditions (\ref{scaleseparation}) underlying the definition of $A_0.$ Therefore, replacing $\lambda_c$ by $(\varepsilon b)$ wherever it appears in $A_0,$ and expanding in $\varepsilon,$ we obtain a series with the coefficient functions independent of $\lambda_c.$ On the other hand, the principal length scale of the rapidly varying part $A_1$ is $\lambda_c,$ whereas the relatively weak dependence on $b$ is expected to be amenable to power expansion in  $1/b$ (increasing $b$ improves conditions (\ref{scaleseparation})). Replacing $1/b \to \varepsilon/\lambda_c$ therein yields a series in $\varepsilon$ whose coefficient functions depend only on $\lambda_c.$ Thus, regarding the coordinate dependence of $A^{(r)}_{0,1},$ it will be assumed in what follows that
$$A^{(r)}_0 = A^{(r)}_0(\bm{x}/b), \quad A^{(r)}_1 = A^{(r)}_1(\bm{x}/\lambda_c)\, \quad r = 0,1,\dots$$
Now, definition (\ref{split}) implies that $\langle A^{(0)}_1 \rangle + \varepsilon\langle A^{(1)}_1 \rangle + \dots = 0,$ and therefore,
\begin{eqnarray}\label{avrapid}
\langle A^{(0)}_1 \rangle = \langle A^{(1)}_1 \rangle = \dots = 0.
\end{eqnarray}
\noindent Within the approximation used throughout this paper, evolution of the flame front envelope is described in terms of $A_0^{(0)}.$ This evolution is affected by the small-scale structure via nonlinear coupling of $A_1$ and $A_0^{(0)}$ in the governing equations.

In subsequent calculations, the natural units are used (see Sec.~\ref{mequation}). This implies, in particular, that $\lambda_c$ is equal numerically to $\varepsilon,$ and that time $t$ is measured in units of $b/U_f.$ We then have the following order-of-magnitude estimates for the functions involved with respect to the parameter $\varepsilon$ ($p$ denotes gas pressure)
\begin{eqnarray}\label{orders0}
\bm{v}_0 &=& O(1), \quad \bm{v}_1 = O(1), \quad p_0 = O(1),
\quad p_1 = O(1),\\
\frac{\partial v_{0i}}{\partial x_k} &=& O(1), \quad
\frac{\partial v_{0i}}{\partial t} = O(1), \quad \frac{\partial
p_0}{\partial x_k} = O(1), \\
\frac{\partial v_{1i}}{\partial x_k} &=&
O\left(\frac{1}{\varepsilon}\right), \quad\frac{\partial
v_{1i}}{\partial t} = O\left(\frac{1}{\varepsilon}\right),
\quad\frac{\partial p_1}{\partial x_k} =
O\left(\frac{1}{\varepsilon}\right),\\
\quad\frac{\partial^2 v_{0i}}{\partial x_k\partial x_l} &=&
O(1)\,, \quad\frac{\partial^2 v_{1i}}{\partial x_k\partial x_l} =
O\left(\frac{1}{\varepsilon^2}\right)\,, \quad i,k,l =
1,2.\label{orders2}
\end{eqnarray}
\noindent Let us proceed to the derivation of effective equations for the functions $\bm{v}_{0}^{(0)}, p_{0}^{(0)}.$

\subsection{Decoupling of dynamical equations}\label{eqdecoupling}

The gas velocity and pressure obey the following bulk equations
\begin{eqnarray}\label{cont}
{\rm div}\,\bm{v} &=& 0\,, \\
\frac{\partial\bm{v}}{\partial t} + (\bm{v}\cdot\bm\nabla)\bm{v} &=&
-\frac{1}{\rho}\bm\nabla p + \bm{g} + \varepsilon
Pr\triangle\bm{v}\,, \label{euler}
\end{eqnarray}
\noindent where $Pr$ is the Prandtl number (the ratio of viscous and thermal
diffusivities, which for gases is of order unity). The role of the viscous term will be discussed in detail later (see Secs.~\ref{ident}, \ref{conclusions}); it is  included already because we are presently concerned with the effect of dissipative processes on the large-scale flame evolution. Extracting the leading $O(1/\varepsilon)$-terms in these equations with the help of estimates (\ref{orders0})--(\ref{orders2}) yields
\begin{eqnarray}\label{flow0}
{\rm div}\,\bm{v}_1^{(0)} &=& 0\,,\\
\frac{\partial\bm{v}_1^{(0)}}{\partial t} +
\left(\left[\bm{v}_0^{(0)} +
\bm{v}_1^{(0)}\right]\cdot\bm\nabla\right)\bm{v}_1^{(0)} &=&
-\frac{1}{\rho}\bm\nabla p_1^{(0)} + \varepsilon
Pr\triangle\bm{v}_1^{(0)}\,, \label{flow0euler}
\end{eqnarray}
\noindent after which collecting $O(1)$ terms gives
\begin{eqnarray}\label{flow1cont}
{\rm div}\,\bm{v}_0^{(0)} &=& 0\,, \\
\frac{\partial\bm{v}_0^{(0)}}{\partial t} +
\varepsilon\frac{\partial\bm{v}_1^{(1)}}{\partial t} +
\left(\left[\bm{v}_0^{(0)} +
\bm{v}_1^{(0)}\right]\cdot\bm\nabla\right)\bm{v}_0^{(0)} &+&
\varepsilon \left(\left[\bm{v}_0^{(0)} +
\bm{v}_1^{(0)}\right]\cdot\bm\nabla\right)\bm{v}_1^{(1)} \nonumber\\
= - \frac{1}{\rho}\bm\nabla p_0^{(0)} &-&
\frac{\varepsilon}{\rho}\bm\nabla p_1^{(1)} + \bm{g} +
\varepsilon^2 Pr\triangle\bm{v}_1^{(1)}\,. \label{flow1euler}
\end{eqnarray}
\noindent Equation (\ref{flow1cont}) already involves only $\bm{v}_0^{(0)},$ while  the other equation couples the slowly and rapidly varying functions. However, the latter disappear upon averaging Eq.~(\ref{flow1euler}). We note, first of all, that
$$\left\langle\frac{\partial\bm{v}_1^{(1)}}{\partial t}\right\rangle = \frac{\partial\left\langle\bm{v}_1^{(1)}\right\rangle}{\partial t} = 0,$$ by virtue of the definitions (\ref{average})--(\ref{avrapid}). Similarly,
$$\left\langle\left(\bm{v}_1^{(0)}\cdot\bm\nabla \right)\bm{v}_0^{(0)}
\right\rangle = \left(\left\langle\bm{v}_1^{(0)}\right\rangle\cdot\bm\nabla \right)\bm{v}_0^{(0)}= 0.$$ Next,
\begin{eqnarray}
\varepsilon\left\langle\bm{\nabla} p_1^{(1)}\right\rangle = \frac{\varepsilon}{L^2}\int\limits_{S}d\bm{s} p_1^{(1)}\,,\nonumber
\end{eqnarray}\noindent where $S$ is the surface of $V,$ $d\bm{s}$ being its element. Since $p^{(1)}_1 = O(1),$ the integral here is $O(L).$ Therefore, according to the choice of $L,$
$$\varepsilon\left\langle\bm{\nabla} p_1^{(1)}\right\rangle = O\left(\frac{\lambda_c}{L}\right) = o(1).$$ The same argument applies to $(\bm{v}_0^{(0)}\cdot\bm\nabla)\bm{v}_1^{(1)},$ as well as to the last term on the right-hand side of
Eq.~(\ref{flow1euler}). Finally, contribution of the last term on its left-hand side also is $o(1).$ Indeed, integrating by parts and taking into account Eq.~(\ref{flow0}), one has
$$\left\langle\left(\bm{v}_1^{(0)}\cdot\bm\nabla\right)
\bm{v}_1^{(1)}\right\rangle = \frac{1}{L^2}\int\limits_{V}d^2\bm{x}~\left(\bm{v}_1^{(0)}\cdot
\bm\nabla\right)\bm{v}_1^{(1)} = \frac{1}{L^2}\int\limits_{S}\left(d\bm{s}\cdot\bm{v}_1^{(0)}\right)\bm{v}_1^{(1)}\,.
$$ In view of Eq.~(\ref{orders0}), the last expression is $O(1/L).$ Hence,
$$\varepsilon\left\langle\left(\bm{v}_1^{(0)}\cdot\bm\nabla\right)
\bm{v}_1^{(1)}\right\rangle = O\left(\frac{\lambda_c}{L}\right) = o(1)\,.$$

Thus, up to $o(1)$-terms, Eq.~(\ref{flow1euler}) reduces upon averaging to the
ordinary Euler equation for the functions $\bm{v}_0^{(0)},$
$p_0^{(0)}$
\begin{eqnarray}\label{aveuler1}
\frac{\partial\bm{v}_0^{(0)}}{\partial t} + \left(\bm{v}_0^{(0)}\cdot
\bm\nabla\right)\bm{v}_0^{(0)} = - \frac{1}{\rho}\bm\nabla
p_0^{(0)} + \bm{g}\,.
\end{eqnarray} \noindent
Thus, to the leading order, the large-scale flow dynamics in the bulk is unaffected by the small-scale wrinkling. This is not unexpected, of course, as the hydrodynamic equations themselves are derived using macroscopic averaging of the more fundamental kinetic equation. That the large-scale flow dynamics turns out to be ideal is just because we average precisely over the distances where the viscous forces are only important. We return to this point in Sec.~\ref{ident}.

The obtained result does not mean that the large-scale flame evolution is completely unaffected by the wrinkling, for this evolution depends also on the local propagation law of the flame front.

\subsection{Renormalization of local propagation law}\label{renorm}

Account of the transport processes inside the flame front modifies the local propagation law. For the sake of simplicity, we shall adopt the simplest form thereof \begin{eqnarray}\label{evolution}&&
(\bm{v}_-\cdot \bm{n}) - \frac{\dot{f}}{N} = 1 -
l_M\frac{f''}{N}\,,
\end{eqnarray}
\noindent where $l_M = O(\lambda_c)$ is the Markstein length \cite{markstein1951}. In the general case, additional terms involving time derivatives appear on the right-hand side of the evolution equation \cite{matalon,class2003}, which require modification of the auxiliary averaging procedure along the front, to be introduced presently. This procedure is needed to reduce Eq.~(\ref{evolution}) to an equation for the slowly varying part of gas velocity. Consider a quantity $A$ defined at the flame front. Given a point $\tilde{x},$ choose $\Delta x =
\Delta x(\tilde{x})$ such that the front length $\EuScript{L}$
between the points $(\tilde{x},f(\tilde{x},t))$ and $(\tilde{x},f(\tilde{x} + \Delta
x,t))$ satisfies
$$\lambda_{\rm c}\ll \EuScript{L} \ll b, \quad \EuScript{L} = O(L).$$
This is always possible since $\EuScript{L}$ is of the order of distance between the two points. Define the average of $A$ over $\{x,y: x \in (\tilde{x},
\tilde{x} + \Delta x), y = f(x,t)\}$ by
\begin{eqnarray}\label{lineaverage}
\langle A \rangle_l = \frac{1}{\EuScript{L}}\int\limits_{\tilde{x}}^{\tilde{x} + \Delta x}dl~A\,, \end{eqnarray}
\noindent where $dl$ is the front line element, $dl = Ndx.$

Using this operation, the unit vector $\bm{n}$ defined at the front, as well as the flame front position itself can be decomposed into two parts corresponding to the scales $b$ and $\lambda_{\rm c}$:
\begin{eqnarray}\label{ntfexpansion}&&
\bm{n} = \bm{n}_0 + \bm{n}_1\,, \quad f = f_0 + f_1\,, \quad
\langle\bm{n}_1\rangle_l = 0\,, \quad \langle f_1\rangle_l = 0\,,
\end{eqnarray}
\noindent which can be further expanded in powers of $\varepsilon,$ as before,
\begin{eqnarray}
\bm{n}_{0,1} = \bm{n}_{0,1}^{(0)} + O(\varepsilon)\,,
\quad f_{0} = f_{0}^{(0)} + O(\varepsilon)\, \quad f_1 =
O(\varepsilon). \label{ntforders}
\end{eqnarray}
\noindent

Let us begin with averaging of the first term on the left-hand side of the evolution equation. In view of Eq.~(\ref{cont}), one can introduce the
stream function $\psi = \psi(x,y,t)$ according to
\begin{eqnarray}\label{vstream}
u = \frac{\partial \psi}{\partial x}\,, \quad w = - \frac{\partial
\psi}{\partial y}\,.
\end{eqnarray}
\noindent Using the operation of bulk averaging
(\ref{average}), the function $\psi$ can be decomposed as
\begin{eqnarray}\label{psidec}
\psi = \psi_0 + \psi_1\,, \quad \langle\psi_1\rangle = 0\,.
\end{eqnarray}
\noindent It follows from Eq.~(\ref{orders0}) that
\begin{eqnarray}\label{psim}
\psi_0 = O(1)\,, \quad \psi_1 = O(\varepsilon)\,.
\end{eqnarray}
\noindent Insertion of the decomposition (\ref{psidec}) into
Eq.~(\ref{vstream}), followed by the bulk averaging gives
\begin{eqnarray}
u_0 = \left\langle\frac{\partial (\psi_0 + \psi_1)}{\partial
x}\right\rangle = \frac{\partial \psi_0}{\partial x} +
\frac{1}{L^2} \left.\int\limits_{\tilde{y}}^{\tilde{y} + L}dy~\psi_1\right|_{\tilde{x}}^{\tilde{x} + L}
= \frac{\partial \psi_0}{\partial x} + O\left(\frac{\lambda_c}{L}\right)\,, \nonumber
\end{eqnarray}
\noindent and a similar equation for $w_0.$ One thus finds, up to $o(1)$-terms, \begin{eqnarray}\label{u0}
u_0 = \frac{\partial \psi_0}{\partial x}\,, \quad w_0 = -
\frac{\partial \psi_0}{\partial z}\,.
\end{eqnarray}
\noindent Next, writing $\bm{n} = (-f'/N,1/N),$ and using Eqs.~(\ref{vstream}), (\ref{psim}) one has for the average normal velocity
\begin{eqnarray}
\left\langle(\bm{v}_- \cdot\bm{n})\right\rangle_l &=&
\frac{1}{\EuScript{L}}\int\limits_{\tilde{x}}^{\tilde{x} + \Delta x}dx~\left(\frac{\partial
\psi}{\partial x} + \frac{\partial\psi}{\partial z}\frac{\partial
f}{\partial x }\right)_- =
\frac{1}{\EuScript{L}}\int\limits_{\tilde{x}}^{\tilde{x} + \Delta x}dx~\frac{d\psi_-}{dx}
= \frac{\psi_{0-}|_{\tilde{x}}^{\tilde{x} + \Delta x}}{\EuScript{L}} + o(1)\,. \nonumber
\end{eqnarray}
\noindent Since $\psi_0$ is a slowly varying function of spatial coordinates, the last expression can be evaluated up to $o(1)$-terms as
\begin{eqnarray}
\psi_{0-}|_{\tilde{x}}^{\tilde{x} + \Delta x} &=&
\psi_{0-}(\tilde{x} + \Delta x, f_0(\tilde{x} +
\Delta x,t),t) - \psi_{0-}(\tilde{x},f_0(\tilde{x},t),t)
\nonumber\\
&=& \left(\frac{\partial\psi_0}{\partial x}
+ \frac{\partial\psi_0}{\partial y}\frac{\partial f_0}{\partial x
}\right)_-\Delta x = \left(u_{0-} -
w_{0-}\frac{\partial f_0}{\partial x}\right)\Delta x\,,\nonumber
\end{eqnarray}
\noindent where Eqs.~(\ref{ntforders}), (\ref{u0}) have been used. Note that
\begin{eqnarray}
n_{x0} = \left\langle - \frac{\partial f/\partial
x}{N}\right\rangle_l = -
\frac{1}{\EuScript{L}}\int\limits_{\tilde{x}}^{\tilde{x} + \Delta x}dx~\left(\frac{\partial f_0}{\partial x} + \frac{\partial f_1 }{\partial x}\right) = -
\frac{\Delta x}{\EuScript{L}}\frac{\partial f_0 }{\partial x} +
o(1)\,,\nonumber
\end{eqnarray}
\noindent and similarly,
\begin{eqnarray}
n_{y0} = \frac{\Delta x}{\EuScript{L}} + o(1)\,.\nonumber
\end{eqnarray}
\noindent Hence, the result of averaging the normal fresh gas velocity along the front takes the form
\begin{eqnarray}
\left\langle(\bm{v}_- \cdot\bm{n})\right\rangle_l = (\bm{v}_{0-} \cdot\bm{n}_0) +
o(1).
\nonumber
\end{eqnarray}
\noindent

It is to be noted that $\bm{n_0}$ is not a unit vector. Up to $o(1)$-terms, its inverse norm is
\begin{eqnarray}\label{veff}
\|\bm{n}_0\|^{-1} = \frac{\EuScript{L}} {\displaystyle\Delta x\sqrt{1 +
\left(\frac{\partial f_0}{\partial x}\right)^2}} \equiv
\beta  \geqslant 1.
\end{eqnarray}
\noindent This quantity has a clear geometrical meaning: $(\beta
- 1)$ represents the relative increase of the flame front length
due to its small-scale wrinkling. In general, the dimensionless quantity $\beta $ is a slowly varying function of $\tilde{x}.$ However, in the case of gravity-driven flame evolution we deal with ($gb/U^2_f\gtrsim 1$), $\beta$ turns out to be position-independent in zero order approximation with respect to $\varepsilon.$ This is because there is no appropriate length parameter to form a dimensionless combination involving $x,$ on which $\beta $ might depend. Indeed, $\lambda_c$ is not a candidate, by the very definition of $\beta .$ On the other hand, since $\beta$ is a geometrical invariant, it is independent of $f^{(0)}_0.$ To prove the latter statement, we note that the coordinate system can always be chosen so that $\partial f_0/\partial x = 0$ at any given point $\tilde{x},$ in which case one can take $\Delta x = L.$ Then $\partial f_0(x,t)/\partial x = O(L)$ for $x\in [\tilde{x},\tilde{x}+L],$ and $\beta $ takes the form
$$\beta  = \frac{\EuScript{L}}{\Delta x} = \frac{1}{L}\int\limits_{\tilde{x}}^{\tilde{x} + L}dx \sqrt{1 + \left(\frac{\partial f^{(0)}_1}{\partial x}\right)^2} + O(L)\,,
$$ which proves the asserted independence of $f^{(0)}_0$ up to terms $o(1).$ Thus, $\beta $ is a functional only of $f^{(0)}_1,$ and hence, is independent of $b$ (Cf. Sec.~\ref{aprocedure}). In other words, $\beta $ does not vary on the length scale $b$ and, {\it ipso facto}, on the smaller scale $U^2_f/g.$

Averaging of the remaining terms in Eq.~(\ref{evolution}) is straightforward. Taking into account Eq.~(\ref{ntforders}), one finds
$$\left\langle \frac{\dot{f}}{N} \right\rangle_l = \frac{1}{\EuScript{L}}\frac{\partial}{\partial t}\int\limits_{\tilde{x}}^{\tilde{x} + \Delta x}dx \{f_0(x,t) + f_1(x,t)\} = \frac{\Delta x}{\EuScript{L}}\dot{f_0}(\tilde{x},t) + O\left(\frac{\lambda_c}{L}\right)\,.$$ Similarly,
$$l_M\left\langle \frac{f''}{N}\right\rangle_l = \frac{l_M}{\EuScript{L}} \int\limits_{\tilde{x}}^{\tilde{x} + \Delta x}dx f'' = \frac{l_M}{\EuScript{L}}f'|_{\tilde{x}}^{\tilde{x} + \Delta x} = O\left(\frac{\lambda_c}{L}\right) = o(1)\,,$$ since $f' = O(1).$ Thus, the evolution equation reduces to
$$(\bm{v}_{0-} \cdot\bm{n}_0) - n_{y0}\dot{f_0} = 1 + o(1).$$ In terms of the zero-order coefficient functions, this can be finally rewritten as
\begin{eqnarray}\label{evoleff}
(\bm{v}^{(0)}_{0-} \cdot\bm{\nu}) - \frac{\dot{f}^{(0)}_0}{\EuScript{N}} = \beta\,,
\end{eqnarray}
\noindent
where $\EuScript{N} = \sqrt{1 + \left(\partial f^{(0)}_0/\partial x\right)^2}\,,$ and $\bm{\nu}$ is the unit vector $\left(-[f^{(0)}_0]'/\EuScript{N},1/\EuScript{N}\right).$ This result shows that from the large-scale point of view, the effect of flame wrinkling amounts to renormalization of the normal flame speed, which in the ordinary units now equals to $\beta U_f.$ That the renormalizing factor $\beta$ is expressed entirely in geometrical terms, with no trace of the parameter $l_M$ characterizing the inner-front properties, is in accord with the idea about compensation of the local effects, mentioned in Sec.~\ref{aprocedure}. The estimates given in \cite{coward1932II} for the front area of 3D flames in wide tubes imply that $\beta \approx 1.4-1.5.$

Finally, it follows from Eqs.~(\ref{aveuler1}), (\ref{evoleff}) that jump conditions at the front for the slowly varying part of gas velocity are the same as in the case of zero-thickness flames. More precisely, if the velocity unit is redefined to be $\beta U_f,$ then the velocity jumps are given by the last two equations (\ref{jumps}) upon replacing $u,w,f\to$ $u^{(0)}_0, w^{(0)}_0,f^{(0)}_0.$ For the sake of notation simplicity, one and the same designation of the flow variables will be used in channels of any width, with the understanding that the above replacement and redefinition of the velocity unit are made when considering flame propagation in wide channels.

\section{Analysis of experimental data}

\subsection{Relevance of 2D picture to flame propagation in tubes}\label{rel}

\begin{figure}
\centering
\includegraphics[width=1\textwidth]{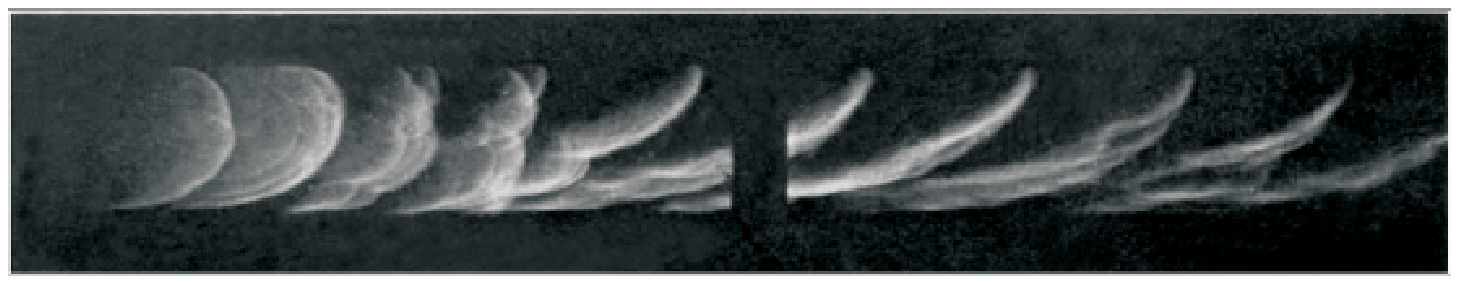}
\caption{}\label{fig5}
\end{figure}

The question as to what extent the results of 2D analysis are applicable in 3D situation is quite nontrivial, and in each particular problem requires special consideration. To settle it in the present case, we turn to Fig.~\ref{fig5} which is a series of snap-shot photographs of the early stage of flame evolution after ignition, made at equal time intervals, in a glass tube of $d=10$\,cm \cite{coward1932I}. It shows that as the flame travels along the tube, its initially hemispherical front not only tilts but also flattens, as is evident from the fact that the front image becomes thinner. This observation suggests that the flame shape in the regime of uniform movement can be described using 2D tools, and implies that the channel width is to be set equal to the tube diameter. As to the speed of uniform movement, it was proposed in \cite{bychkov1997,bychkov2000} that the 2D values are to be multiplied by a factor $\approx 1.5$ to obtain the flame speed in a tube (of diameter equal to the channel width), the reason being that this relation holds for bubbles of similar shape. However, as was already mentioned in Introduction, the theory of bubble motion is irrelevant to flame propagation. As a matter of fact, there is a solid reason to expect that no such factor is to be introduced in flame theory. The point is that unlike bubbles, flames propagate because they consume fuel; the rate of this consumption is proportional to the flame front area, which is determined by the slope of front elements, rather than their azimuth with respect to the tube axis. Thus, the relative velocity increase of 2D and 3D flames is the same if they have the same average slope.

The misleading analogy with bubbles helps to consider this issue from a somewhat different standpoint. To get insight into how flame dimensionality might affect its speed, compare propagation of flames and bubbles in a channel and in a tube of equal size. To simplify  geometry, it is more convenient to deal with vertical propagation, wherein 3D shapes can be assumed axisymmetric. Consider first bubble motion, and let the 3D bubble shape be obtained from the corresponding 2D channel pattern by rotating it around the tube axis. The bubble speed depends essentially on the relative cross-sectional area of the jet formed between the bubble and the walls, which is evidently larger in the 3D case. The same geometrical reason suggests that if it were possible to obtain an axisymmetric flame by rotating a 2D pattern around the tube axis, then the speed of this flame would indeed be larger than the speed of its 2D original as proposed in \cite{bychkov1997,bychkov2000}, since the front slope grows with distance from the tube axis. However, such rotation would violate flow continuity, for the gas elements burnt near the tube wall come from regions of lesser radial distance from the axis (Cf. Fig.~\ref{fig6}), whereas the flow far upstream remains homogeneous. Therefore, instead of increasing the speed of propagation, 3D flame has to reduce its longitudinal spread to maintain flow continuity. On the other hand, retaining continuity in the case of bubbles of fixed shape is not a problem, for their motion is not bound to satisfy such condition as the constancy of local consumption rate. We thus see that the difference in the mechanisms governing propagation of bubbles and flames implies essentially different sensitivity to the flow dimensionality.

It can be added that in complete analogy with the 2D case, the effect of small-scale 3D front structures amounts to renormalization of the local propagation law (the proof is almost identical to that given in the preceding section). To be sure, these arguments do not affirm equality of propagation speeds of 2D and 3D flames, but they point toward certain rigidity of the flame speed with respect to manifestations of the three-dimensional nature of the problem, such as occurrence of pronounced 3D front structures, or circularity of the tube cross-section. On these grounds, the speed of uniform movement of flame in a tube of diameter $d$ will be identified henceforth with the speed of steady 2D flame propagating in a channel of equal width, $b=d.$

\subsection{Assessment of heat losses}\label{losses}

For flat-on-average flames, heat losses to the tube walls are usually of minor importance in regard to the flame speed, for their relative value rapidly decreases with increasing tube diameter. Thus, although preventing flame propagation in very narrow tubes, heat losses are practically negligible in tubes of a few centimeters diameter. For instance, the speed of downward propagation of planar methane-air flames near the limits of inflammability (in which case the gravity and transport effects completely stabilize planar front) is the same in tubes of $d=5$\,cm and $d=23$\,cm   \cite{mason1920}, implying that even for slowest flames the cooling effect is negligible already in tubes of $5$\,cm diameter. Things are different, however, for horizontally propagating flames, because the large flame spread along the tube considerably enhances heat outflow from the front, reducing thereby the flame speed. The rate of this outflow is, in the ordinary units,
$$\bar{q} = K\Sigma \Delta T,$$ where $\Sigma \approx \pi d \times (U/\beta U_f)d/2$ is the area of the tube surface surrounding the front (the factor $1/2$ accounts for the fact that heat is transferred only from the hot side of the front), $\Delta T$ the temperature difference between the cold wall and burnt gas, and $K$ the heat transfer coefficient. On the other hand, the longitudinal heat flux due to deflagration is
$$q = \rho Q \times U \times \pi d^2/4,$$ where $Q$ is the specific heat of combustion, and $\rho$ is the cold gas density. The dimensionless ratio
\begin{eqnarray}\label{loss}
\delta \equiv \frac{\bar{q}}{q} = \frac{2K}{\rho c_p \beta U_f}\,,
\end{eqnarray}
\noindent
where $c_p\equiv Q/\Delta T$ is the average specific heat capacity of fresh gas, gives the relative velocity drop due to the heat losses to the walls. It is seen that this parameter depends neither on $d,$ nor on $U.$ In application to flame propagation in methane-air mixtures under standard conditions, $\rho\approx 1.2$\,kg/m$^3,$ $c_p$ ranges 1.3 to 1.8\,kJ/kg$\cdot$K, depending on the methane concentration, while the heat transfer coefficient for tubular air/metal heat exchangers ranges 5 to 35\,W/m$^2\cdot$K \cite{hexchange}. As no relevant information is given in the cited works\footnote{That the heat losses to the tube walls can significantly affect flame propagation was recognized already by Mallard and Le Chatelier who estimated them for flat flames and concluded that the effect is negligible in sufficiently wide tubes, the critical tube diameter being approximately $5$\,cm (Cf.~\cite{mason1917}). This conclusion was apparently shared by Coward and Hartwell despite their discovery that the large increase of flame speed in horizontal tubes is the result of large longitudinal flame spread.} to further specify this parameter, the mean value $K=20\,$W/m$^2\cdot$K will be used in all subsequent estimations of $\delta$. Substitution of the above figures into Eq.~(\ref{loss}) shows that the velocity drop can be as large as $30\%$ (this value is attained near the limits of inflammability).

\subsection{Identification of physical solutions}\label{ident}

The family of all possible steady flame patterns was described in Sec.~\ref{numeric}. To select practically relevant solutions, one has to invoke properties of the physical gas flow existing far downstream. The point is that despite ideality of the large-scale flow, proved in Sec.~\ref{eqdecoupling}, its asymptotic structure downstream is essentially determined by the gas viscosity. Indeed, even if viscosity is so small that its effect is negligible at finite distances from the front, it leads to gradual decay of vorticity produced in the flame front, whereby the flow becomes homogeneous sufficiently far downstream (for simplicity, we discard viscous drag exerted by the walls, which results in the logarithmic velocity profile). It is worth recalling in this connection that the master equation (\ref{master2}) itself was derived in the sense of analytic continuation to the limit of zero viscosity \cite{jerk4}.

\begin{figure}
\centering
\includegraphics[width=1\textwidth]{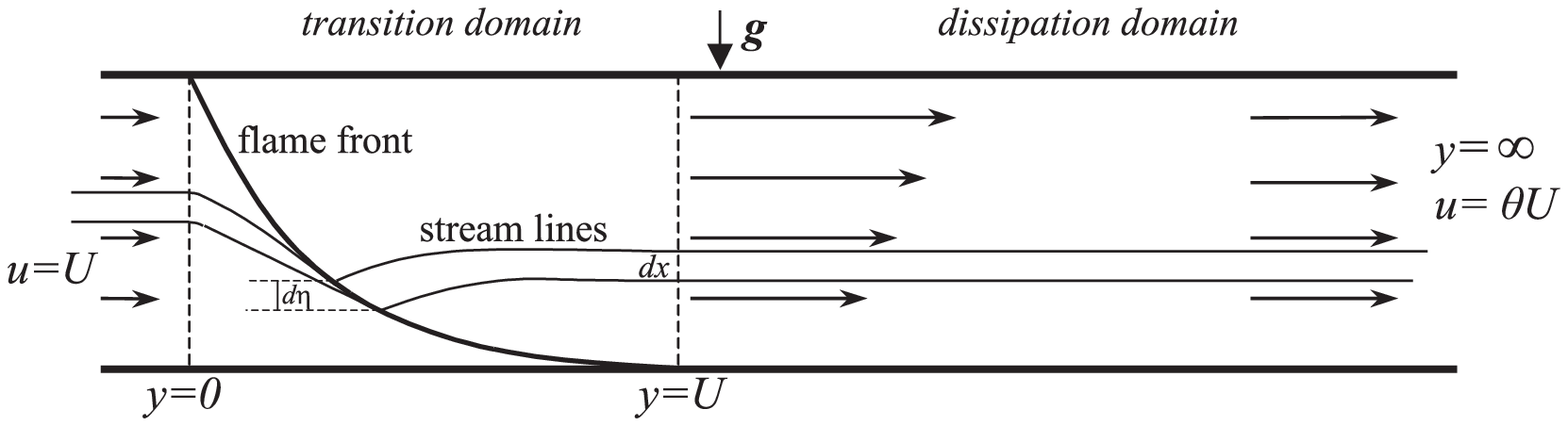}
\caption{}\label{fig6}
\end{figure}

To identify relevant solutions, we note that the vorticity decay is accompanied by a loss of kinetic energy of the burnt gases. Hence, criterion for selecting physical solutions follows from the energy conservation law: the total loss of kinetic energy downstream must be compensated by its gain in the process of combustion, {\it i.e.,} by the gravitational work done on gases in the transition domain.\footnote{In the leading order of the large-slope expansion, the kinetic energy increase due to velocity jump across the front is negligible in comparison with the gravitational work.} The latter is the region where transversal component of the gas velocity is nonzero (the gravitational power vanishes together with $w$). The reasoning used to derive Eq.~(\ref{bcondition3}) implies that this is the region indented in Fig.~\ref{fig6} by two vertical broken lines. The above criterion can be expressed in terms of the kinetic energy flux through a vertical section at the point $y,$ to be denoted $q_T(y)$: the total kinetic energy loss downstream is the difference  $q_T(\infty) - q_T(U),$ whereas the kinetic energy gain is $q_T(U) - q_T(0).$ Equating the two expressions yields
\begin{eqnarray}\label{criterion}
q_T(U) = \frac{1}{2}\left\{q_T(\infty) + q_T(0)\right\}\,.
\end{eqnarray}
\noindent By virtue of the total mass conservation, velocity of the asymptotic homogeneous flow is, in the natural units, $\theta U$ (recall that according to our definition of the velocity unit in wide channels, this is $\theta U \beta U_f,$ in the ordinary units). Hence,
\begin{eqnarray}\label{criterionr}
\frac{1}{2}\left\{q_T(\infty) + q_T(0)\right\} = \frac{1}{2}\left\{\theta U\frac{(\theta U)^2}{2\theta} + U\frac{U^2}{2} \right\} = \frac{(\theta^2 + 1)U^3}{4}\,.
\end{eqnarray}
\noindent
To express the left-hand side of Eq.~(\ref{criterion}) in terms of the fresh gas velocity, we note that the flow continuity and Bernoulli integral imply the following relations between the on-shell burnt gas velocity and pressure, and their values at the line $y=U$:
\begin{eqnarray}
\theta N(\eta)d\eta &=& u(x,U)dx, \nonumber \\
\frac{u^2_+(\eta)}{2} + \theta p_+(\eta) - g\eta &=& \frac{u^2(x,U)}{2} + \theta p(x,U) - gx,\nonumber
\end{eqnarray}
\noindent where $(x,U)$ is the intersection of the line $y=U$ with the stream line that crosses the front at $(\eta,f(\eta)),$ Fig.~\ref{fig6}. Since $w(x,U) = 0,$ the combination $[\theta p(x,U) - gx]$ is independent of $x,$ so that it can be replaced by $[\theta p_+(1) - g].$ Taking into account also that $p_+(1) = p_-(1) - \alpha,$ and using Bernoulli integral for the upstream flow, we find
\begin{eqnarray}
q_T(U) &=& \int\limits_{0}^{1}dx u(x,U)\frac{u^2(x,U)}{2\theta} \nonumber\\ &=& \int\limits_{0}^{1}d\eta f'(\eta)\left[\frac{u^2_-(1) - U^2}{2} - \frac{\theta -1}{2\theta}u^2_-(\eta) + \frac{\theta - 1}{\theta}g(\eta - 1)\right].\nonumber
\end{eqnarray}
\noindent With the help of the relations $f(1) = U,$ $f' = u_- - (1-x)u'_-,$ this can be finally written as
\begin{eqnarray}\label{criterionl}
q_T(U) = \frac{U}{2}\left[u^2_-(1) - U^2\right]  + \frac{\theta -1}{\theta}\int\limits_{0}^{1}d\eta \left[u_- - (1 - \eta)u'_-\right]\left[ - \frac{u^2_-}{2} + g(\eta - 1)\right].
\end{eqnarray}\noindent The right-hand sides of Eqs.~(\ref{criterionr}), (\ref{criterionl}) are functions of $U.$ Physical solutions correspond to intersections of their graphs, as illustrated by Fig.~\ref{fig7} in the case $\theta = 7,$ $g=61.3.$ The left solid curve is the plot of $q_T(U)$ for Type I solutions $U \in (4.45,4.753),$ whereas the right solid curve represents the initial part of $q_T(U)$ for Type II solutions $U > 6.615$; the dotted curve is the plot of $(\theta^2 + 1)U^3/4.$ It is seen that there is one physical Type I solution with $U = 4.72,$ and one physical Type II solution with $U = 6.96.$ Numerical scrutiny of the equation for $u_-$ has shown that this situation is generic: For every set of parameters $\theta, g,$ there is exactly one physical solution of each type.

To conclude this section, let us return to the issue touched upon at the end of Sec.~\ref{mequation}, namely anticipated vanishing of the last term in Eq.~(\ref{master2}). As was already noted in Sec.~\ref{reduction}, its exact value cannot be found using the functions we work with, because they are obtained as solutions to Eqs.~(\ref{evolutiongen1}), (\ref{3rel}), (\ref{master5}) continued over the narrow region near $x=0$ where the large-slope approximation is not valid. Despite its narrowness, this region gives rise to a non-negligible contribution to the integral (\ref{gamma}), for the integrand is proportional to the derivative of gas velocity which rapidly varies near the wall. Under such circumstances, making use of approximate values cannot decide whether $\gamma$ vanishes on the physical solutions, but it gives valuable qualitative information about its behavior on the family of all solutions. Specifically, combining Eqs.~(\ref{jumps}), (\ref{vorticity}), the constant $\gamma$ can be written as
$$\gamma = - \alpha\int\limits_{0}^{1}d\eta w_+\left\{\frac{d\ln v_+}{d\eta} - \frac{g}{v^2_+}\right\}.$$ Taking into account that $w_+\sim \alpha,$ this expression shows that in general, $\gamma\sim \alpha^2.$ For instance, in the above example $\gamma = 52.4$ for $U = 4.6$ (Type I solution), and $\gamma = 32.1$ for $U=8.0$ (Type II solution). However, the approximate value of $\gamma$ significantly drops down on the physical solutions: its values for Type I and Type II physical solutions are $9.2$ and $13.5,$ respectively. Moreover, $\gamma$ has a root in a vicinity of the physical solution: $\gamma=0$ for $U=4.735$ and $U=6.76.$ This observation gives support to the hypothesis that the exact $\gamma$ vanishes on physical solutions. If true, this would deliver an intrinsic criterion for selecting physical solutions, without the need to invoke asymptotic properties of the flow downstream.

\subsubsection{The range of applicability of the theory. Accuracy assessment}\label{assessment}

The asymptotic formula (\ref{hcurvedf1}) for $\EuScript{H}$-operator requires for its validity only that the front slope be large, $f'\gg 1,$ or equivalently, $U\gg 1,$ and so evidently does Eq.~(\ref{evolutiongen1}). The relative value of neglected terms in these formulas is $1/U^2.$ However, derivation of Eqs.~(\ref{3rel}), (\ref{master5}) essentially uses the estimates $u=O(U),$ $w=O(1)$ to omit $w$ in comparison with $u.$ Taking into account that the magnitude of $w$ is $\alpha$ (since $w_0 = \alpha/2$), this implies a stronger condition $U \gg \alpha.$ Switching to the ordinary units and recalling that $U\sim \sqrt{g d}\,,$ we thus obtain the formal criterion of applicability of the theory
$$\frac{\alpha U_f}{\sqrt{g d}} \ll 1.$$
The meaning of this condition is that the ratio on its left-hand side estimates the relative error introduced by the above-mentioned omission.

Even without detailed analysis of solutions, it is clear that flame propagation with $U<\alpha/2$ is not covered by the present theory, since Eqs.~(\ref{bcondition3}), (\ref{bcondition4}) imply that in this case $u_-(0)<w_-(0),$ and so the relative error exceeds unity, at least at the flame head. Hence, solutions obtained in Sec.~\ref{largeslope} accurately describe the front position and velocity profiles, only if $U\gg\alpha/2.$ As to determination of the physical value of $U$ itself, the requirement is not actually that stringent, for $U$ is an integral quantity. The dominant contribution to $U$ comes from the region where the front slope is largest, that is, from the lower part of the channel. In Type II solutions, $u_-(x)$ increases with $x,$  $u'_-$ being proportional to $\alpha,$ while $w_-(x)$ monotonically decreases, so that condition $u_-\gg w_-$ can be considered practically satisfied, provided that $u_-\gtrsim w_-$ at the upper wall, {\it i.e.} if
\begin{eqnarray}\label{range}
U \gtrsim \frac{\alpha}{2}\,.
\end{eqnarray}\noindent In Type I solutions, the $u$-component of gas velocity behaves as before, while the transversal component first increases, picks at $x\approx 0.2,$ where it is generally of order $u_-,$ and then rapidly decreases. Since the pick is in the upper part of the channel, criterion (\ref{range}) can be applied to this type of solutions as well. Accuracy of the theoretical value $U$ within the range (\ref{range}) can now be estimated as follows. Let $r$ denote the relative error in $U.$ This error comes  mainly from the region where $w/u \gtrsim r.$ On the other hand, since $U$ is equal to the front length, $r$ is approximately the length of the front arc belonging to this region, relative to the total front length. Therefore, $r$ can be determined as the ordinate of intersection of the curves $w/u$ and $f/U.$ An example is given in Fig.~\ref{fig8} for Type I solution with $\theta=5,$ $b=40$\,cm, $U_f = 10$\,cm/s, and $U=11.3,$ in which case $r\approx 10\%.$ The relative error is smaller in wider channels and for slower flames, being about $20\%$ at the boundary of the domain defined by (\ref{range}). As subsequent analysis shows, larger errors outside this domain make comparison of the theory and experiment inconclusive.

\subsection{Comparison with experimental data on methane-air flames}\label{comparison}

We now go over to a detailed comparison of the obtained results with the experimental data on methane-air flames, given in \cite{coward1932I,coward1932II}. The flame speed data to be analyzed is collected in figures 2 and 3 of \cite{coward1932I}, the first of which is reproduced in Fig.~\ref{fig9}. The experimental marks are due to Mason and Wheeler (open circles) \cite{mason1917}, Mason (triangles) \cite{mason1923}, Coward and Greenwald (crosses) \cite{coward1928}, and Coward and Hartwell (filled circles) \cite{coward1932I}. The curves represent an attempt to interpolate the experimental data for various methane concentrations.
Inflections in the curves in the region $10$\,cm$<d<20$\,cm indicate the onset of formation of complex front structures observed in wide tubes.

To begin with, we note a very important evidence contained in the table reproduced in Fig.~\ref{fig9}, namely that the experimental data for the speed of uniform movement (denoted S.U.M. in the table) cannot be approximated by a {\it single} function of the form
\begin{eqnarray}\label{naive}
U = C(\theta)\sqrt{gd}\,,
\end{eqnarray}\noindent suggested by the naive dimensional analysis.
To be more specific, consider a pair of flames in a tube of $20$\,cm diameter, corresponding to $c=8.02\%$ and $c=12.05\%.$ According to the table, they propagate with the speeds $U = 102$\,cm/s and $U = 84$\, cm/s, respectively. On the other hand, both flames have $\theta \approx 6.9$ \cite{calc}. Furthermore, the heat loss parameter $\delta$ also has the same value $0.07$ in the two instances. Therefore, according to the naive formula, the flames ought to travel with the same speed. A similar pair for $d=10$\,cm is $c=8.21\%$ and $c=11.94\%,$ in which case $\theta = 6.94,$ and $\delta=0.064,$ but the corresponding flame speeds are $84$\,cm/s and $68$\,cm/s. That the differences in the observed values are well above experimental errors is clearly seen from the same table which contains data for pairs of close concentrations at the points being compared. For instance, the above figure for $c=8.02\%$ comes together with that for $c=8.15\%$ ($U = 103$\,cm/s), whereas the one for $c=12.05\%$ -- with $c=12.04\%$ ($U = 81$\,cm/s). Apparently, the authors of \cite{coward1932I} tabulated the data so carefully because they recognized the difficulty of fitting the flame speed data into ansatz (\ref{naive}), though did not state this explicitly.\footnote{These authors emphasize that the relation between flame speed and tube diameter cannot be expressed in the form $U=Cd^k$ with constant $C,k,$ but relate this to the existence of inflections in the curves $U(d).$}

Directly related to this is a peculiarity exhibited by the flame speed considered as a function of methane concentration, $U(c),$ namely, existence of inflection points at $c\approx 8\%$ and $c\approx 11.4\%.$ It is seen from the data in Fig.~\ref{fig9} that after a relatively moderate rise between $c = 6\%$ and $c \approx 7.2\%,$ $U(c)$ increases significantly more rapidly on the interval $c\in (7.2\%,9\%),$ after which it remains almost constant for $c$ up to $10.4\%.$ Similarly, the fall of $U(c)$ on the interval $c \in (11\%,12.2\%)$ is considerably larger than on $(12.2\%,13.5\%).$ The ratio of the flame speed at the central plateau to its values at the foothills ({\it i.e.,} at $c\approx 7.2\%$ and $c\approx 12.2\%$) exceeds $1.5.$ This behavior cannot be explained by differences in the normal flame speed at different fuel concentrations, for $U$ is independent of $U_f$ when $gd > U^2_f.$

Proceeding to explanation of the observed flame behavior, let us first identify the factors that determine variation of the flame speed with fuel concentration. They include

(1) Dependence of the flame speed on the gas expansion coefficient,

(2) Heat losses to the tube walls,

(3) Existence of two distinct regimes of steady flame propagation.

\noindent Regarding (1), numerical analysis of the differential equation for $u_-(x)$ shows that the flame speed of physical Type I (Type II) solutions increases (decreases) as $\theta$ increases, though this dependence is relatively weak. Next, heat losses significantly affect propagation of flames near the inflammability limits, giving rise only to a few-percent correction to the flame speed in near-stoichiometric mixtures. At last, the existence of two types of solutions with the flame speed ratio about $1.5$ is the key factor for explaining the aforementioned peculiarities in the flame speed data. As will be shown below, the inflections in $U(c)$ correspond to transitions between Type I and Type II solutions. The qualitative picture of the dependence $U(c)$ is thus as follows. Mixtures near the limits of inflammability and near-stoichiometric mixtures sustain respectively Type I and Type II regimes of flame propagation. The joint effect of factors (1),(2) explains the moderate rise of the flame speed on the interval $c\in (6\%,7.2\%),$ and also its moderate fall on $(12.2\%,14\%).$ In near-stoichiometric mixtures, on the contrary, these factors counteract partially compensating each other, which produces the plateau in $U(c)$ for $c\in (9\%,10.4\%).$ Transitions between the two regimes explain the rapid changes of the flame speed in mixtures with $c\in (7.2\%,9\%)$ and $c \in (11\%,12.2\%).$

To carry out quantitative comparison of the theory and experiment requires some auxiliary experimental data. The normal velocities deduced in \cite{coward1932II} from the front area calculations are somewhat lower than the planar flame speeds, because of discard of the finite front-thickness effects. The actual values of $U_f$ will be adopted from \cite{law1992,liao2004}. The gas expansion coefficients are based on the adiabatic flame temperature calculated with the help of \cite{calc}. The heat losses are taken into account by multiplying the theoretical value $U$ by $(1-\delta),$ where $\delta$ is given by Eq.~(\ref{loss}). In this formula, $\rho$ is set equal to $1.2$\,kg/m$^3$ (neglecting its variation with methane concentration) and $K=20\,$W/m$^2\cdot$K in all cases considered below. The quantity $c_p$ is calculated in each instance using the data provided in \cite{calc}. Regarding $\beta,$ the estimates given in \cite{coward1932II} imply that $\beta\approx 1.45$ in a tube of $24\,$cm diameter. In the absence of more detailed information, $\beta=1.45$ will be used in all wide tubes ($d\geqslant 20\,$cm). In tubes of moderate width, $\beta = 1,$ by definition.

Let us start from the lower inflammability limit represented in Fig.~\ref{fig9} by the lowest curve interpolating experimental data for $c=6\%.$ The following is a detailed example of the computational procedure used below to compare the theory and experiment at various methane concentrations. Recalling that the stoichiometric methane/air composition is at $c=9.5\%,$ one first finds the equivalence ratio for the mixture under consideration: $\phi = 0.61.$ It is then used to determine adiabatic flame temperature with the help of \cite{calc}, which for  standard conditions (1\,atm, 300\,K) yields $1680$\,K. Hence, $\Delta T = 1380$\,K and $\theta = 5.64.$ Type I and Type II physical solutions found numerically for this $\theta$ read $$U_I = 18.2\sqrt{d}\,, \quad U_{II} = 29.3\sqrt{d}\,,$$ where $d$ is to be expressed in centimeters. Next, to take into account the heat losses, one uses $0.9$\,MJ/mol for methane's heat of combustion, together with the found $\Delta T,$ to compute $c_p\approx 1.3$\,kJ/kg$\cdot$K. The normal flame speed $U_f = 11$\,cm/s is inferred from Fig.~2 of \cite{liao2004}. Substitution into Eq.~(\ref{loss}) gives $\delta = 0.16$ and $\delta = 0.23$ for tube diameters $d\geqslant 20$\,cm and $d\leqslant 10$\,cm, respectively. Finally, condition (\ref{range}) implies that the above numerical solutions are valid as far as $U\gtrsim 2.3,$ or in the ordinary units, $U \gtrsim 2.3\times 11$\,cm/s$\ \approx 25$\,cm/s in tubes of moderate width, and $U\gtrsim 2.3\times 1.45\times 11$\,cm/s$\ \approx 40$\,cm/s in wide tubes. It is seen from Fig.~\ref{fig9} that for the given methane concentration, the range of applicability of the theory covers all available experimental data. The computed flame speeds for Type I and Type II solutions, combined with this data, are plotted in Fig.~\ref{fig10}(a). Clearly, flames propagating in a $6\%$ mixture favor regime described by Type I solutions (solid curve) in tubes of any diameter. The same is true of $7\%$ mixture, Fig.~\ref{fig10}(b), where $U_f = 20$\,cm/s, $\theta = 6.26.$ In this case, condition (\ref{range}) requires $U\gtrsim 53$\,cm/s, hence only results for $d\geqslant 10$\,cm are shown. On the contrary, burning in mixtures close to the stoichiometric composition is of Type II, as is evident from Fig.~\ref{fig11} representing $U(d)$ for $9\%$ and $10\%$ mixtures: within the error of calculation, all experimental marks in these two cases belong to the curve representing Type II solutions (dotted line). The computed values of the main parameters used to draw the curves for various mixtures are collected in Table I, together with the relevant parameters of physical solutions. The latter include the constants $U(d)/\sqrt{d}$ for each type of solutions, and the error of calculation determined as described in Sec.~\ref{assessment}. Given in the last two columns are pairs of reference values of the error -- $r_m$ for $d=10$\,cm, and $r_w$ for $d=100$\,cm. The figures for the heat loss parameter $\delta$ refer to wide tubes; in tubes of $d\leqslant 10$\,cm they are to be multiplied by $1.45.$ Regarding mixtures with $c=8\%$ to $12\%,$ condition (\ref{range}) implies that the theoretical results are amenable to verification only in wide tubes. Taking $c=10\%$ for instance, one has $U_f = 43$\,cm/s, $\theta = 7.5,$ and $U_I=1.4,$ $U_{II} = 2.03$ in a tube of $10$\,cm diameter, whereas $\alpha/2 = 3.25;$ to put it in terms of accuracy, the error of calculation $r=23\%$ makes comparison of the theory and experiment inconclusive in this case.

The above consideration leads us to the conclusion that the rapid change in flame speed, observed in the range of concentrations $c=7.2\%$ to $c=9\%,$ represents transition between Type I and Type II regimes of flame propagation. To get insight into the nature of this transition, let us turn to Fig.~\ref{fig12} containing the flame speed data in $8\%$ mixture. The following consideration shows that this data points toward existence of an intermittency of the two regimes in transient mixtures. Namely, it is seen that while the experimental marks for $d=30.5$\,cm are clearly of Type II, the mark for $d=96.5$\,cm belongs to neither Type I nor Type II solution. In fact, the error of calculation for this tube diameter is $r=12\%$ for Type I and $r=10\%$ for Type II solution, so the distance of the mark from either curve is well above the error of calculation (this mark is due to \cite{mason1917} according to which the experimental error is about $2$--$3\%$).\footnote{As to the mark for $d=20$\,cm, the data is inconclusive, since the two curves are close to each other, whereas calculational error is relatively large for this tube diameter, $r=20\%.$} This observation can be easily explained if we admit that the flame undergoes transitions between the two regimes as it travels in the tube. This explanation is suggested by the very fact that the given mixture is transient, and is most naturally formulated in terms of stability of the solutions. Indeed, that the mixtures with $c\lesssim 7\%$ propagate only Type I flames means evidently that Type II solutions are unstable in these mixtures (or, at least, that they are considerably less stable than those of Type I), and that situation is opposite in mixtures with $c\gtrsim 9\%.$ It is therefore natural to expect that in the $8\%$ mixture, the two types of solutions are approximately even regarding their stability properties. Intermittency occurs if both solutions are unstable or neutrally stable.

The coexistence of two solutions with similar stability properties is not sufficient to establish intermittency in the case when they are stable, for it would not exist without a mechanism that continuously provokes switching of the propagation regimes. However, such mechanism is readily identified. Namely, it is not difficult to see that fluctuations of the heat flux due to heat losses into the walls are sufficient to trigger transition from one regime of flame propagation to the other. For this to happen, the heat inflow/outflow, $\Delta Q,$ necessary to accelerate/decelerate the flame must be comparable to the difference of kinetic energies of gases in the two regimes, and occur on a time interval less than the characteristic time of perturbation damping. For the gravity-controlled flame evolution this time is of order $U_f/g.$ Multiplying it by $\bar{q}$ from Sec.~\ref{losses} yields $4\times 10^3$\,W for Type I solution ($U = 4.25$), or in the natural units, $5\times 10^4;$ for Type II solution ($U=6.3$) the figure is $1.5$ times larger. On the other hand, the kinetic energy of gases in the volume surrounded by $\Sigma$ can be evaluated as $$E_K(U) \approx \frac{(\theta U)^2}{2\theta}\,\times \frac{\pi U}{4}\,,$$ which gives $E_K(U_I) \approx 2\times 10^2$ and $E_K(U_{II})\approx 7\times 10^2.$ Therefore, even a few-percent local fluctuation of the heat transfer coefficient will produce $\Delta Q$ sufficient to switch the propagation regime. The resulting intermittency can in principle be observed as a recurring appearance of an inflection point at the front head (Cf. Figs.~\ref{fig2}, \ref{fig3}). The above stability consideration implies that flames in transient mixtures are particularly susceptible to such fluctuations, but the described mechanism may work also in mixtures where the two regimes of flame propagation are of markedly different stability. In that case, a more intensive fluctuation of the heat flux is required to perturb the flame, as the damping time is shorter in more stable regimes, and no proper intermittency occurs.

For an experimental evidence of the proposed scenario, let us turn to the work \cite{coward1928} remarkable, in particular, by its scrupulous exposition of the experimental procedures. Regarding flame propagation in a tube $30$\,m long and $30.5$\,cm in diameter, the authors mention that ``some experiments demonstrated that there was a varying of the flame speed throughout the tube, an occurrence which was more frequent the more disturbed the external atmosphere,'' and emphasize that experiments were made only when the wind was negligible near the open end of the tube (Cf. pp. 17, 20, and 22 in \cite{coward1928}). The latter precaution excludes direct atmospheric impact on the burnt gas flow, implying that the observed variations in the flame speed are caused by variations of the external conditions along the tube. That the value of the heat transfer coefficient strongly depends on the ambient air conditions is well known.

Returning to consideration of the function $U(c),$ one finds that the flame speed behavior at methane concentrations larger than $10.4\%$ is quite similar to that at $c<10.4\%.$ Namely, comparison of the theoretical and experimental data for a $11\%$ mixture, Fig.~\ref{fig13}(a), shows that this mixture sustains Type II regime, whereas flames propagating in $13\%$ and $14\%$ mixtures are of Type I in tubes of any diameter, as is evident from Figs.~\ref{fig13}(c),\ref{fig13}(d), and as was contemplated in the qualitative analysis of the experimental data, given earlier. Figure \ref{fig13}(b) suggests that the $12\%$ mixture is transient, again in conformity with the observation that the flame speed decreases most rapidly on the interval $c\in (11\%,12.2\%),$ although proximity to the right end-point of this interval does not give good grounds to affirm intermittency in this mixture: the distance of the experimental marks for $d=30.5$\,cm and $d=96.5$\,cm from the curve representing Type I solution only slightly exceeds the calculational errors ($r = 16\%$ and $12\%,$ respectively).

\section{Discussion and conclusions}\label{conclusions}

Analysis of the horizontal flame propagation, carried out in Sec.~\ref{onshell}, revealed existence of two different types of steady flame patterns formed under the strong influence of the gravitational field. Characteristic of Type I flames is the existence of an inflection point at the flame front head, with an associated pronounced maximum of transversal component of the on-shell fresh gas velocity. The longitudinal component of the gas velocity increases monotonically along the front in both cases, but more rapidly in Type II flames which always propagate faster than Type I flames under the same conditions. Next, it was found in Sec.~\ref{ident} that there is exactly one solution of each type, satisfying the physical requirement of asymptotic homogeneity of the burnt gas flow. At last, detailed comparison with the experimental data on methane-air flames, performed in Sec.~\ref{comparison}, showed that both regimes are realized in practice: mixtures near the limits of inflammability propagate only Type I flames, whereas flames in mixtures close to the stoichiometric composition follow the faster Type II regime. This result resolves the difficulty mentioned in Sec.~\ref{comparison}, namely that the experimental data for the flame speed versus tube diameter cannot be fitted into a single-valued function: Two flames with equal $\theta$ have different velocities in tubes of the same diameter because they propagate in regimes of different type. Furthermore,  inflection points in the functions $U(c)$ correspond to transitions between the two regimes of flame propagation, which take place in the regions characterized by rapid changes of the flame speed.

The comparison also clarifies the origin of infections in the curves representing flame speed versus tube diameter. Namely, the small-scale wrinkling that develops in wide tubes effectively increases the normal flame speed, reducing thereby the heat losses: according to Eq.~(\ref{loss}), they are damped by a factor $\beta = 1.45.$ This effect is especially pronounced in mixtures near inflammability limits where the heat losses are significant, and is clearly seen in Figs.~\ref{fig13}(c),\ref{fig13}(d). Of course, this leaves open the question why complex front structures develop in tubes of $d\gtrsim 10$\,cm, but this is entirely a stability issue. Incidentally, occurrence of these structures cannot be described as a result of development of the Darrieus-Landau instability: Behavior of flame perturbations in the highly nonlinear flows generated by flames propagating with $U \gg U_f$ is completely different from that found for planar flames \cite{jerk3}. In this connection, a general fact to be emphasized is that formation of steady flame patterns as well as dynamics of disturbances of flames with $U \gg U_f$ is essentially determined by vorticity production in the flame. In the considered case of horizontal flame propagation, this process is controlled by the baroclinic effect.

\begin{figure}
\centering
\includegraphics[width=0.6\textwidth]{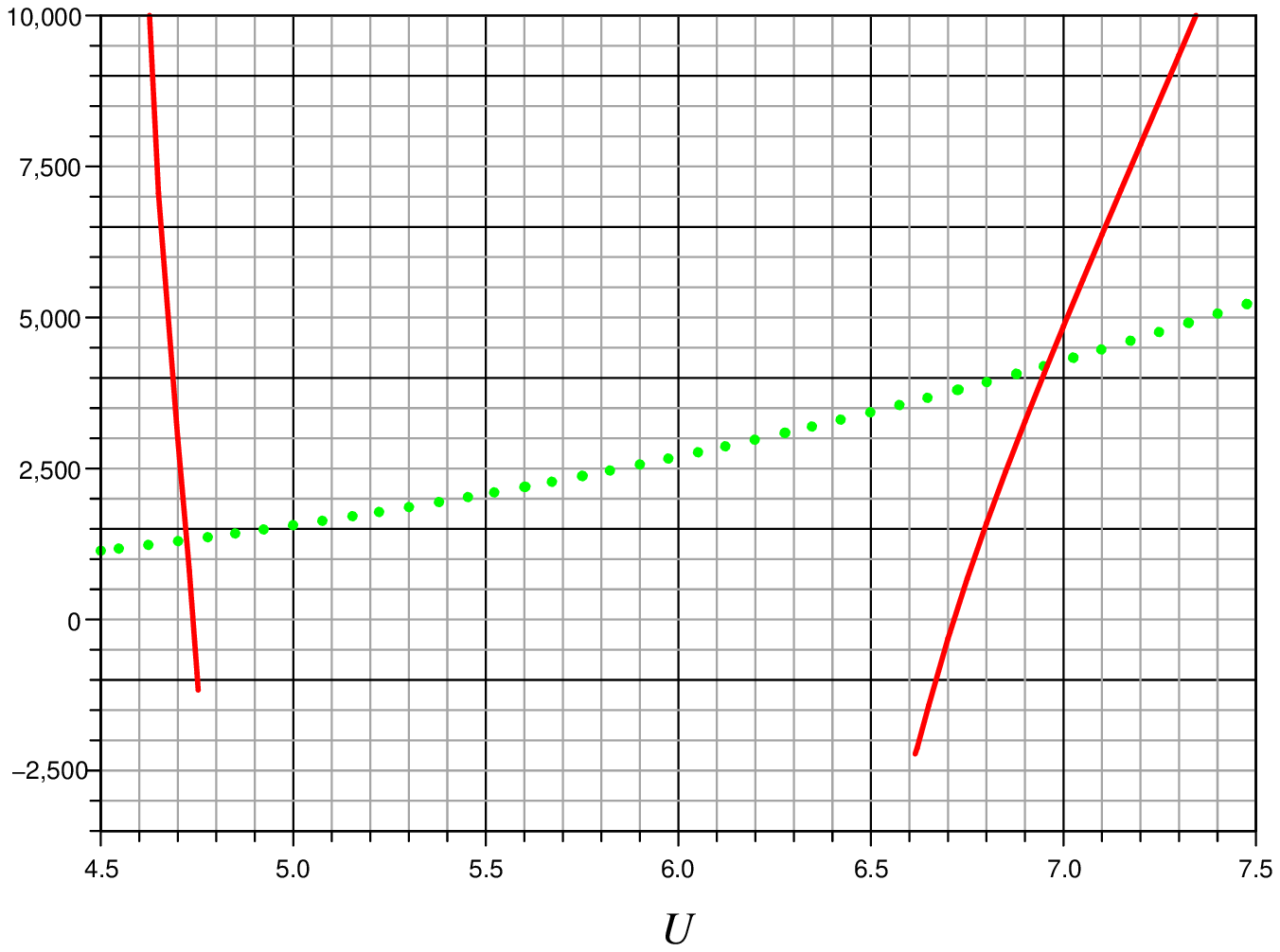}
\caption{}\label{fig7}
\end{figure}

\begin{figure}
\centering
\includegraphics[width=0.4\textwidth]{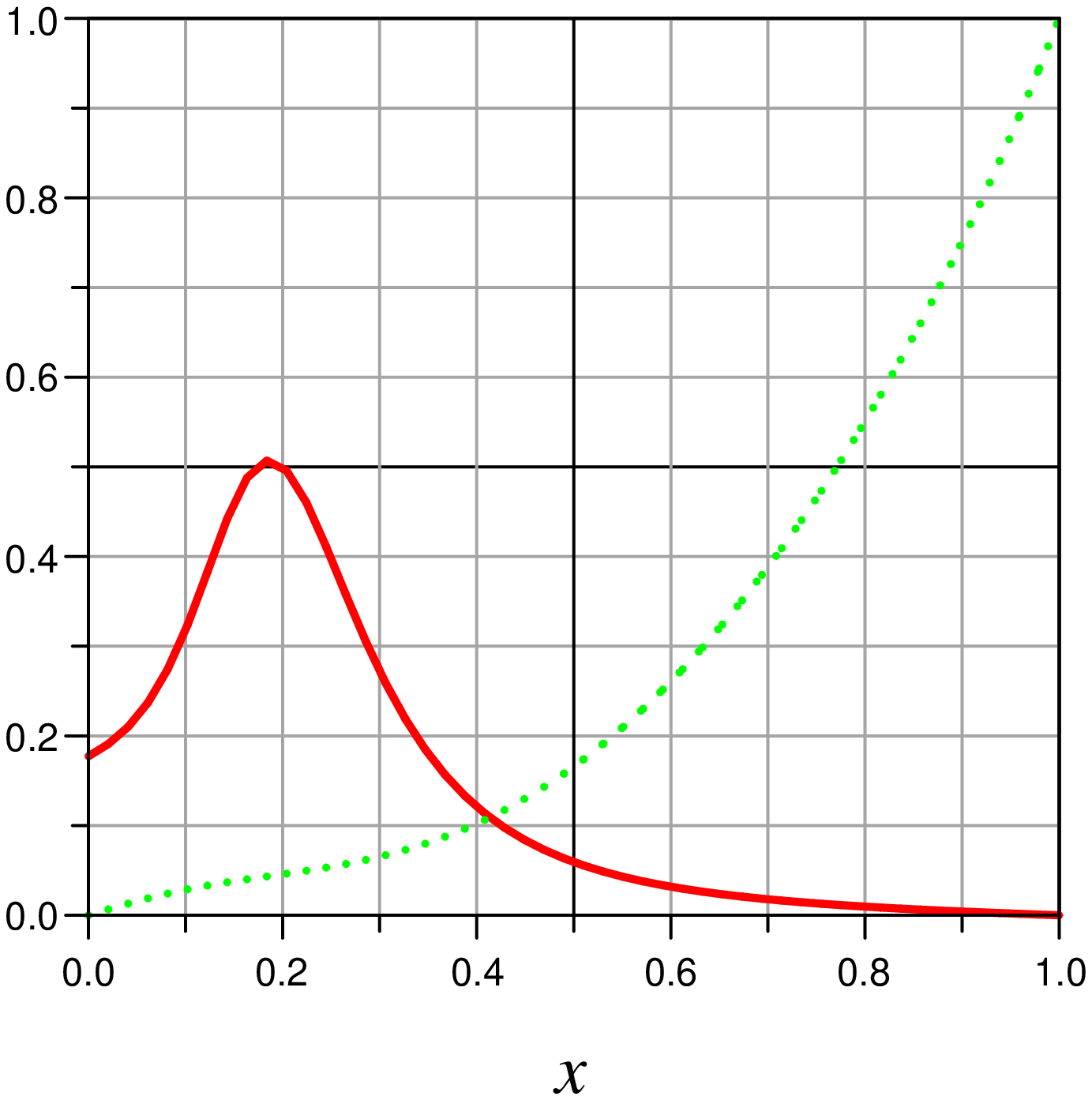}
\caption{}\label{fig8}
\end{figure}

\begin{figure}
\centering
\includegraphics[width=0.8\textwidth]{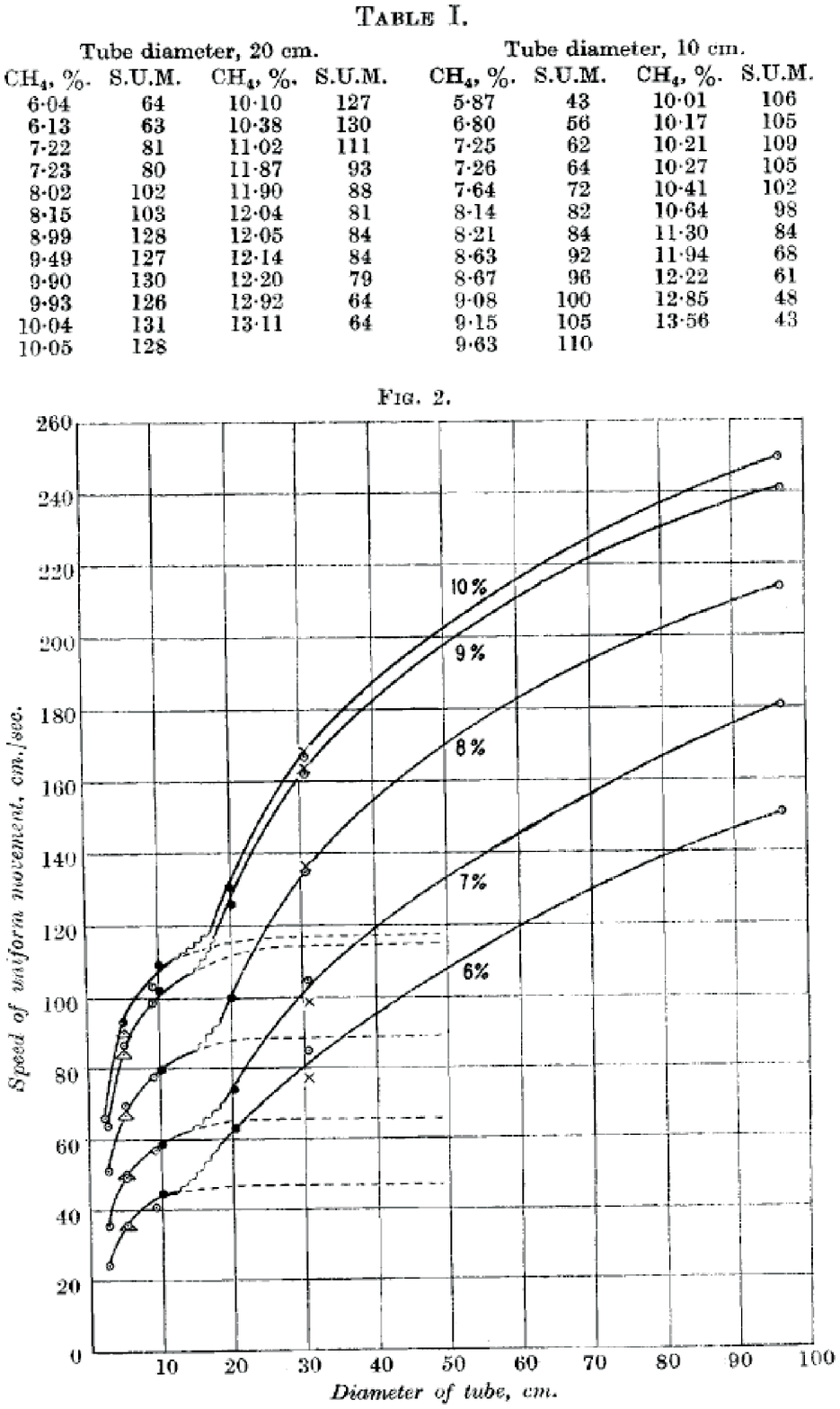}
\caption{}\label{fig9}
\end{figure}

\begin{figure}
\includegraphics[width=0.46\textwidth]{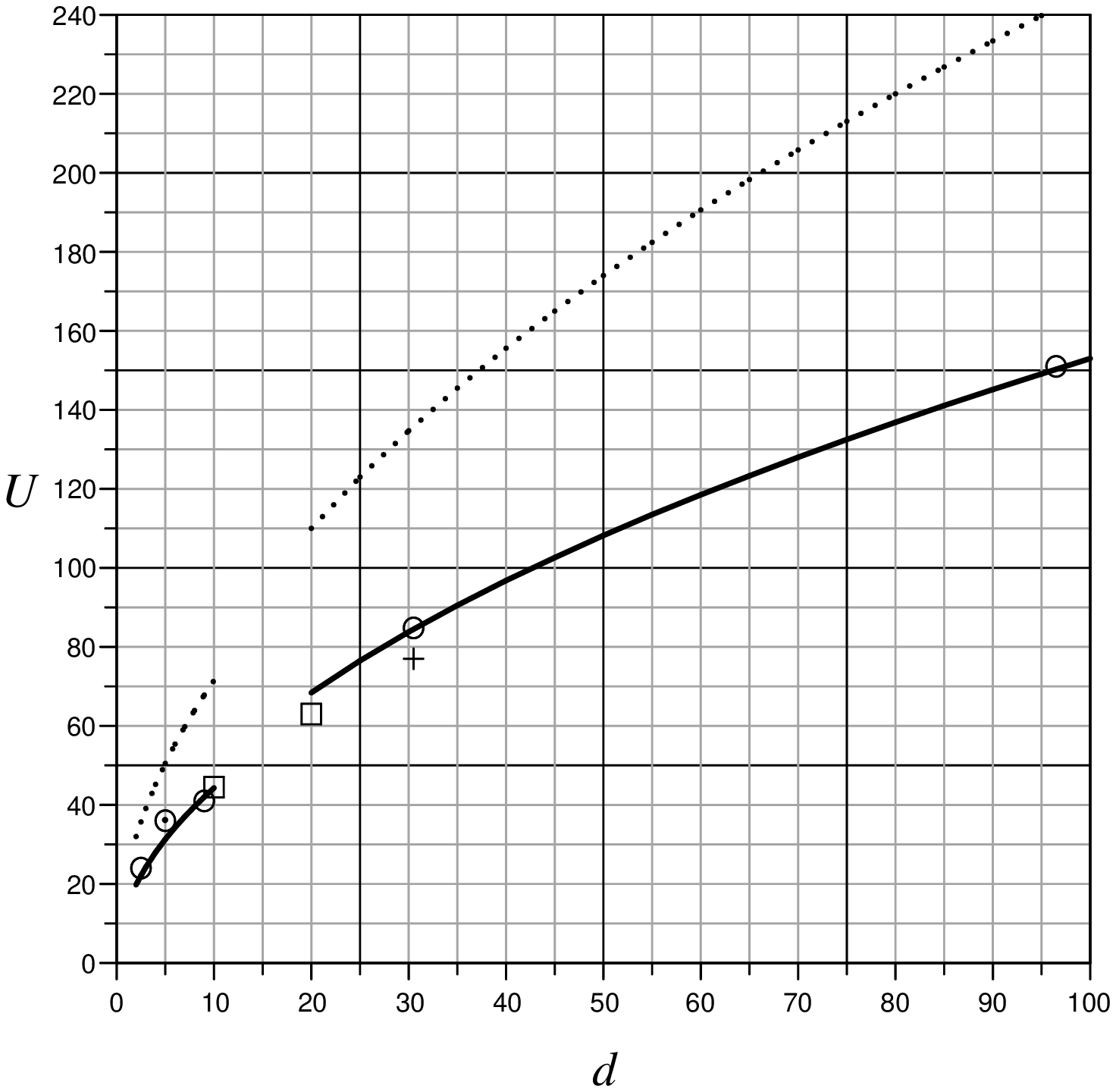}\hspace{1cm}
\includegraphics[width=0.46\textwidth]{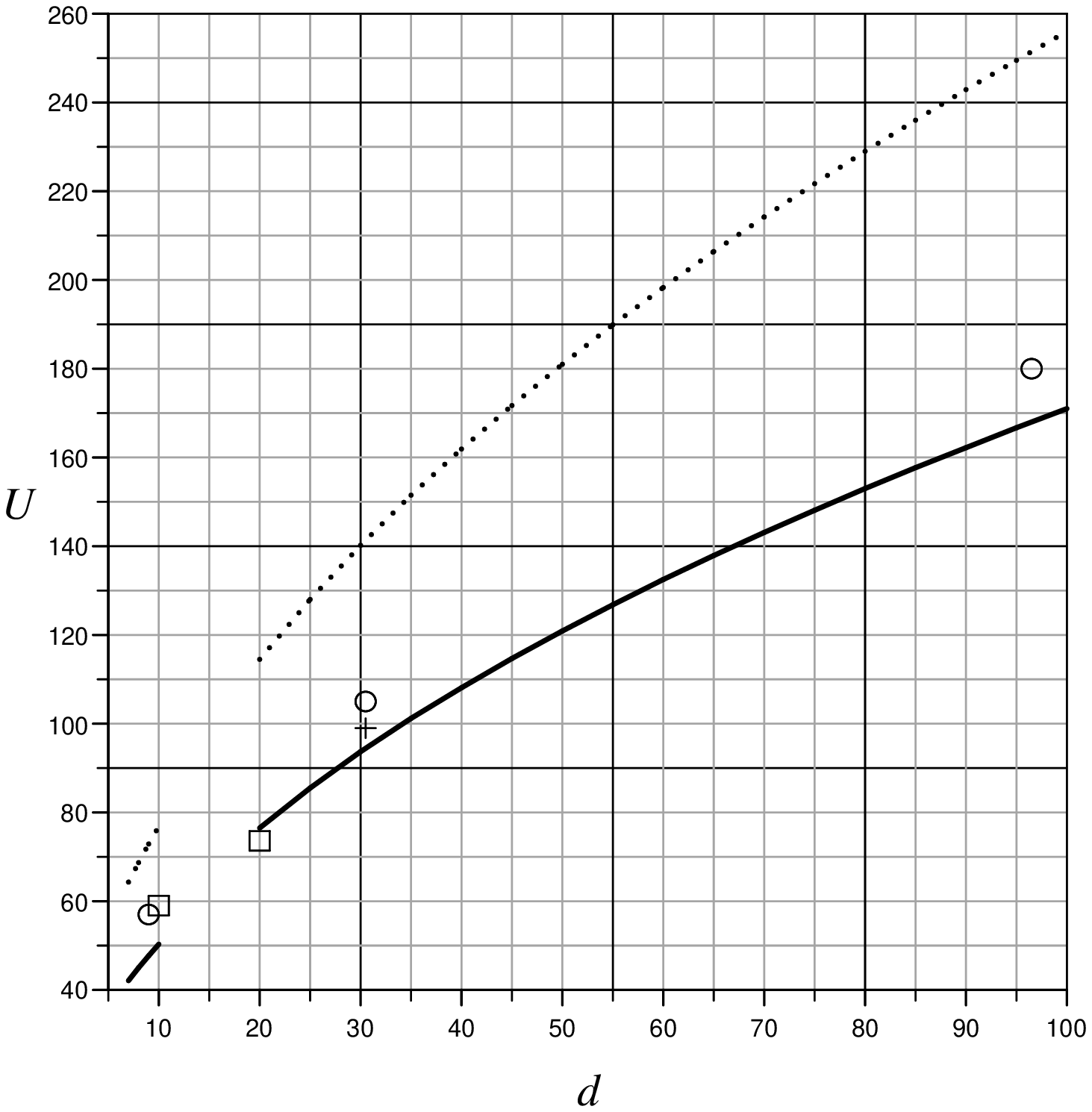}\\
\hspace{0.8cm}(a)\hspace{8.2cm} (b)
\caption{}\label{fig10}
\end{figure}

\begin{figure}
\includegraphics[width=0.46\textwidth]{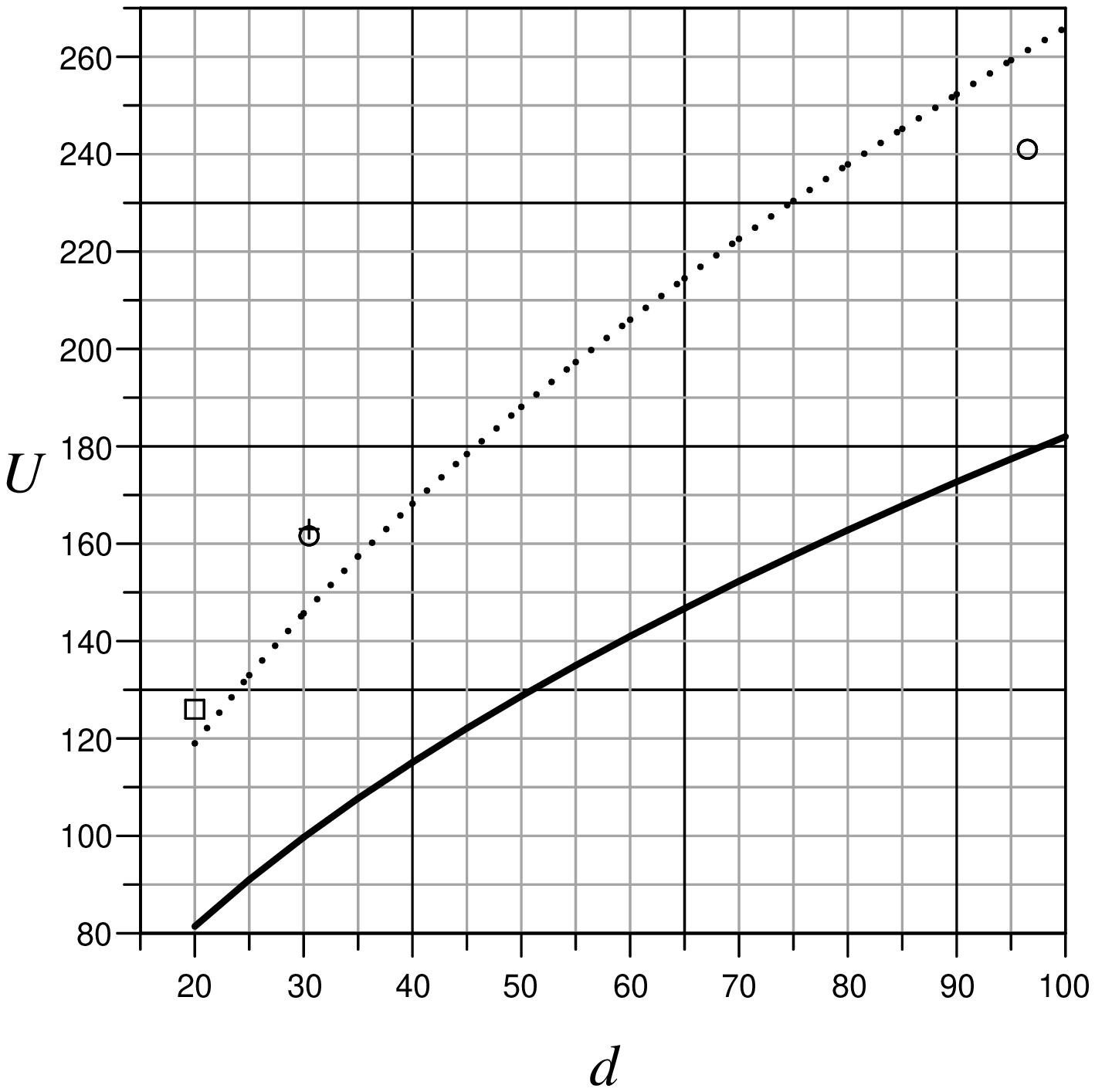}\hspace{1cm}
\includegraphics[width=0.46\textwidth]{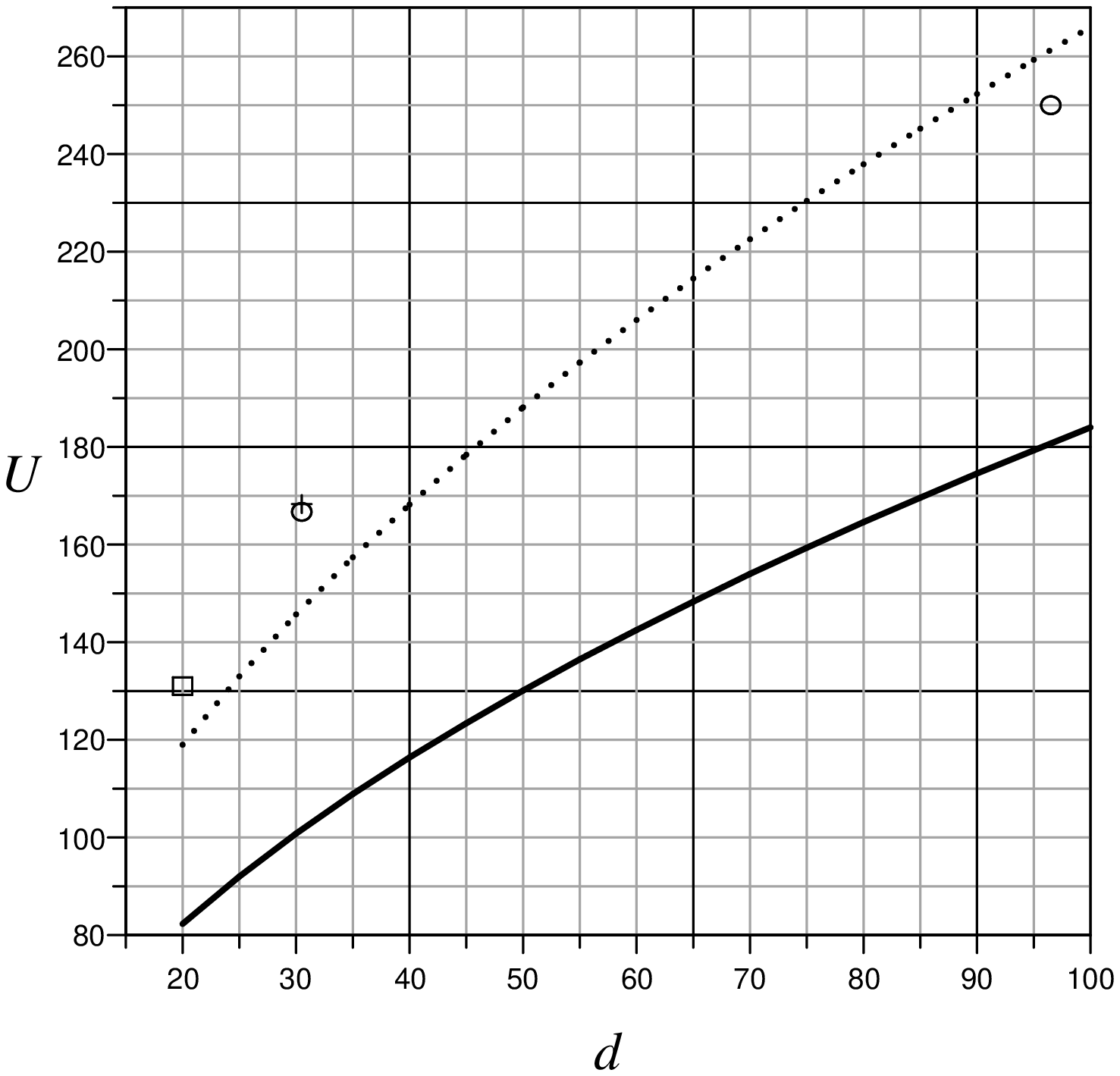}\\
\hspace{0.8cm}(a)\hspace{8.2cm} (b)
\caption{}\label{fig11}
\end{figure}

\begin{figure}
\centering
\includegraphics[width=0.46\textwidth]{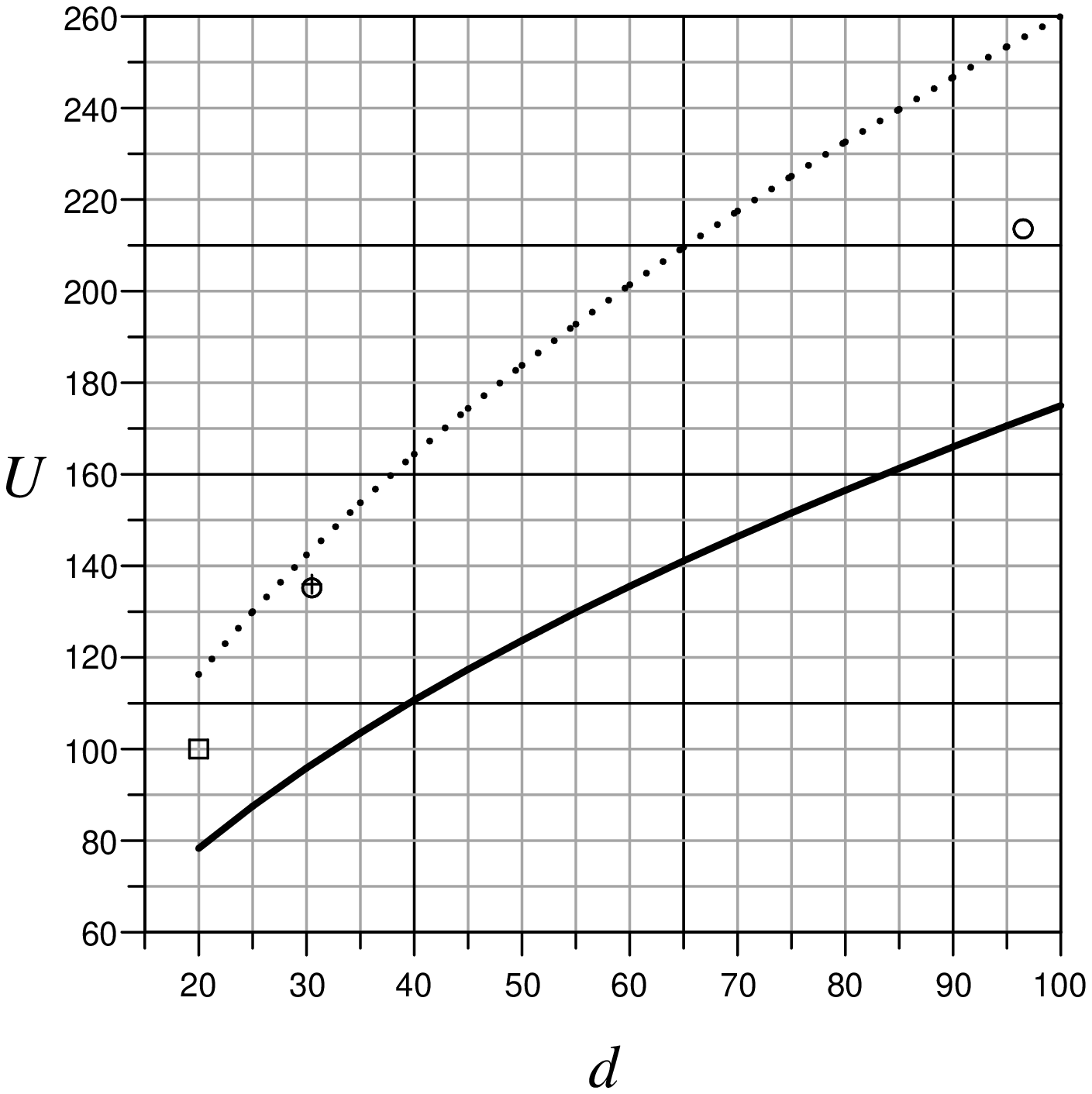}
\caption{}\label{fig12}
\end{figure}

\begin{figure}
\includegraphics[width=0.46\textwidth]{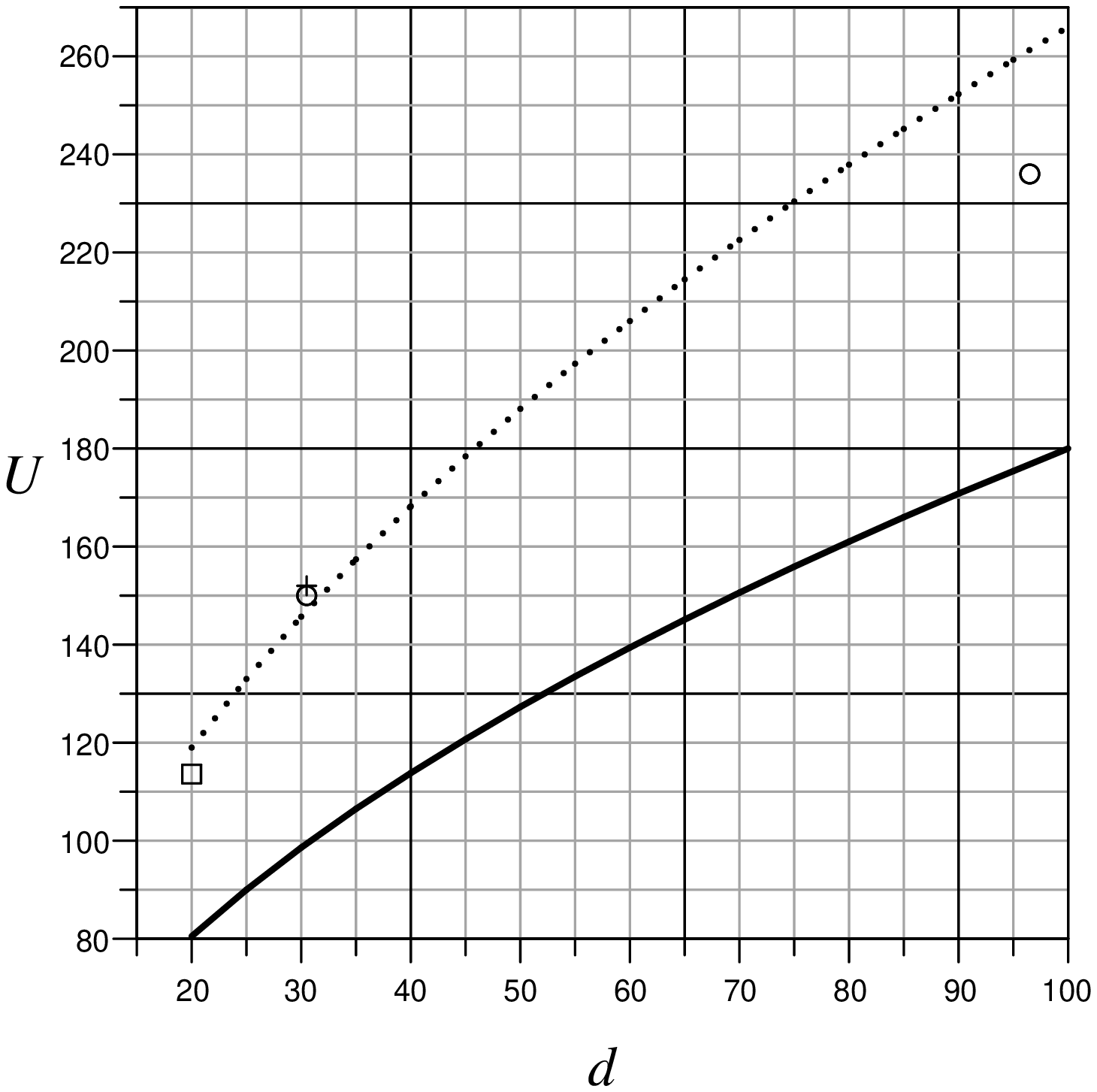}\hspace{1cm}
\includegraphics[width=0.46\textwidth]{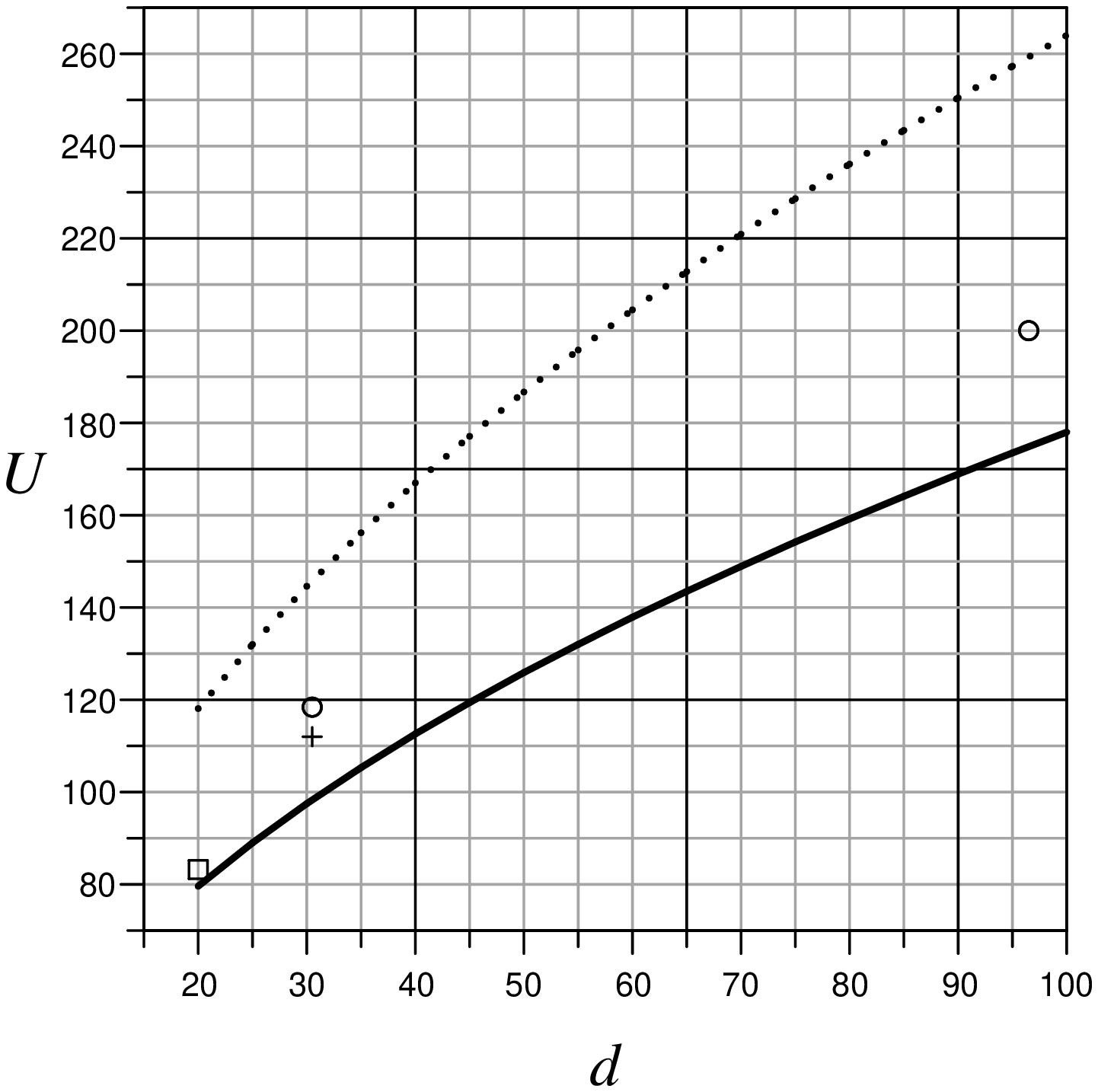}\\
\hspace{0.8cm}(a)\hspace{8.2cm} (b)\\\vskip2cm
\includegraphics[width=0.46\textwidth]{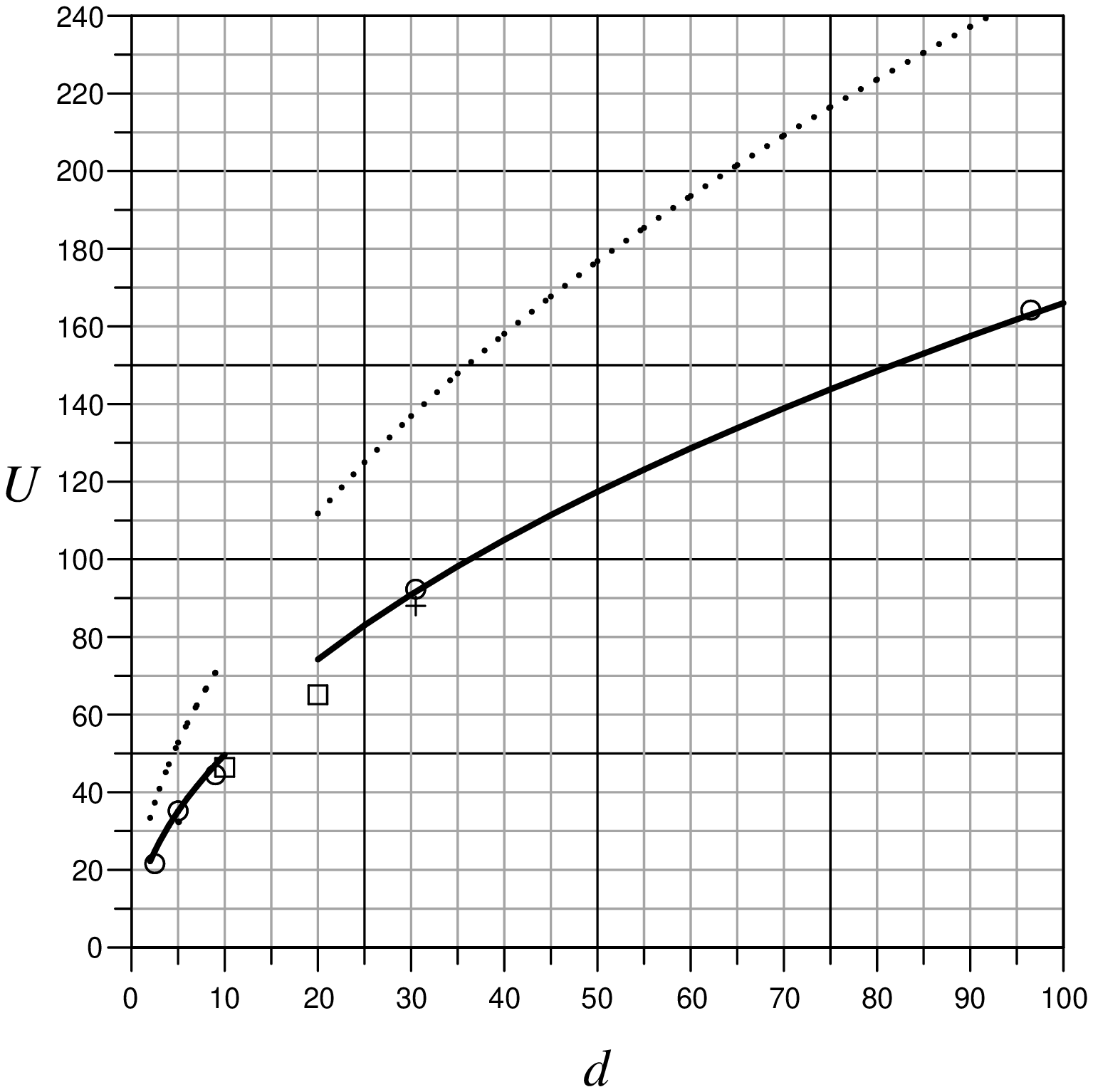}\hspace{1cm}
\includegraphics[width=0.46\textwidth]{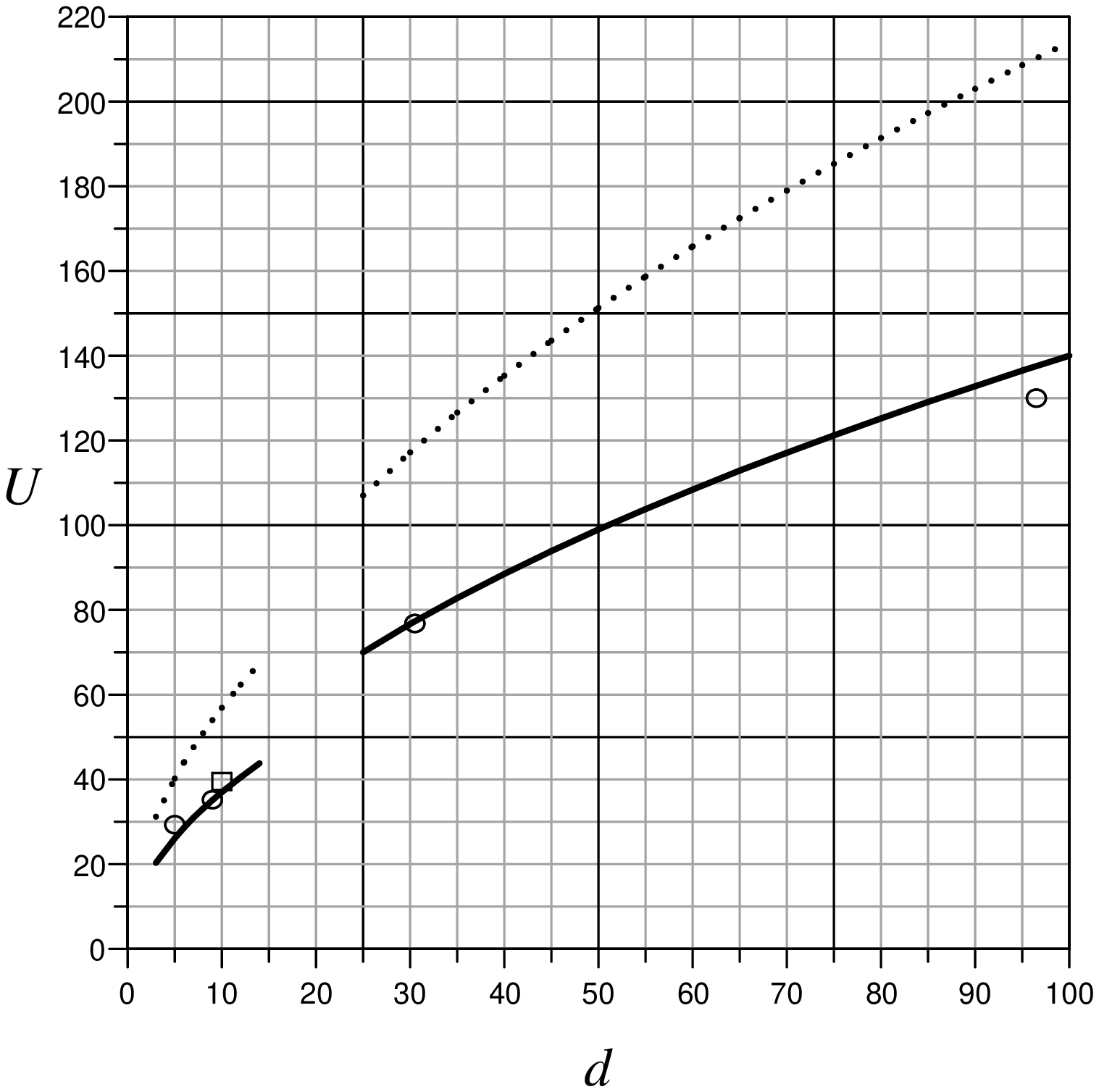}\\
\hspace{0.8cm}(c)\hspace{8.2cm} (d)
\caption{}\label{fig13}
\end{figure}

A technical issue that deserves special discussion is the identification of physical solutions. As was already mentioned in Sec.~\ref{ident}, the requirement of asymptotic homogeneity of the burnt gas flow, used to select physical solutions, is intimately related to the fact that within the on-shell approach, flow dynamics is considered in the limit of vanishing viscosity. It was also pointed out that the criterion identifying physical solutions may probably be formulated as the requirement of vanishing of the parameter $\gamma$ that describes effective gravity in the system, although this cannot be asserted based on approximate evaluation of $\gamma$ in Sec.~\ref{ident}. What can be affirmed, however, is that $\gamma$ is certainly nonzero far from physical solutions, and in view of equivalence of gravity and translatory acceleration, this has an important implication regarding the phenomenon of flame self-acceleration. To explain the point, let us allow for small but nonzero viscosity. This introduces no appreciable changes in the structure of Type I or Type II solutions, but significantly changes criterion for selecting physical solutions. The reason is that the viscous drag exerted by the walls, which now cannot be neglected, causes additional energy losses by the gas flow, increasing with the distance traveled by the flame. For sufficiently small viscosity (such that the viscous length $\gtrsim b U$) this process can be considered quasi-steady. To compensate for the additional losses, a larger kinetic energy flux $q_T(U)$ is required. An immediate consequence is that this causes {\it deceleration} of Type I flames, and {\it acceleration} of Type II flames, as is evident from Fig.~\ref{fig7}. Since Type II solutions exist for all $U>U_2,$ flame acceleration continues until compressibility effects come into play. This observation opens a way to describe the flame pre-acceleration which is well known to be an essential (and least studied theoretically) stage of deflagration-to-detonation transitions.

It is to be noted, finally, that the methods developed in this paper can be used almost directly to study high-speed regimes of vertical flame propagation.

\acknowledgments{I thank Dr. O.~Peil who raised my interest to the problem of horizontal flame propagation by providing me with a copy of \cite{coward1932I,coward1932II}.}

\newpage

\begin{table}
\begin{tabular}{c||cccc|cccc}
\hline\hline
  \hspace{0,2cm} $c,$ \hspace{0,2cm}
  & \hspace{0,6cm} $\theta$ \hspace{0,6cm}
  & \hspace{0,3cm} $U_f,$ \hspace{0,3cm}
  & $c_p,$
  & \hspace{0,7cm}$\delta$ \hspace{0,7cm}
  & \hspace{0,1cm} $U_I/\sqrt{d},$ \hspace{0,1cm}
  & \hspace{0,1cm} $U_{II}/\sqrt{d},$ \hspace{0,1cm}
  & $r_m,$
  & $r_w,$
  \\
  \hspace{0,2cm} $\%$ \hspace{0,2cm}
  &
  & \hspace{0,3cm} cm/s \hspace{0,3cm}
  & kJ/kg$\cdot$K
  &
  & $\sqrt{cm}/s$
  & $\sqrt{cm}/s$
  & \hspace{0,2cm} $\%$ \hspace{0,2cm}
  & \hspace{0,2cm} $\%$ \hspace{0,2cm}
  \\
\hline\hline
6 & 5.64 & 11 & 1.33 & 0.16 & 18.2 & 29.3 & 15 & 9\\
7 & 6.26 & 20 & 1.37 & 0.08 & 18.6 & 28.5 & 18 & 11\\
8 & 6.85 & 30 & 1.42 & 0.05 & 18.8 & 27.9 & 22 &  13\\
9 & 7.31 & 37 & 1.47 & 0.04 & 19.0 & 27.7 & 21 & 12\\
10 & 7.48 & 43 & 1.52 & 0.03 & 19.1 & 27.6 & 23 & 13\\
11 & 7.23 & 38 & 1.57 & 0.04 & 19.0 & 27.7 & 22 & 12\\
12 & 6.90 & 26 & 1.64 & 0.05 & 18.9 & 27.9 & 20 & 12 \\
13 & 6.59 & 12 & 1.71 & 0.11 & 18.7 & 28.1 & 14 & 8 \\
14 & 6.27 & 5 & 1.79 & 0.25 & 18.6 & 28.5 & 10 & 6 \\
\end{tabular}
\caption{Parameters used to draw theoretical curves in Figs.~10--13.} \label{table1}
\end{table}

~\newpage

~\\\\
\centerline{\large\bf Figure captions}\\\\

Fig.1: Flame propagation in horizontal channel of width $b.$ (a) Symmetric pattern obtained by reflection of the physical channel with respect to $y$-axis. Vertical arrows depict acceleration of gravity, $\bm{g},$ which in the image domain turns out to be directed upward, leading to a discontinuity in $\bm{g}$ at $x=0.$ (b) Physical realization of the discontinuity by a gravitating layer placed at $x=0.$ \dotfill 24

Fig.2: Front position and on-shell fresh gas velocity of Type I flames with $\theta = 5$ (solid line) and $\theta =4.5$ (dotted line), propagating in a channel of $b=40$\,cm; normal flame speed $U_f=10$\,cm/s, dimensionless speed of uniform movement $U=11.28.$ \dotfill 25

Fig.3: Same for Type II flames with $\theta =7.5$ (solid line) and $\theta=5$ (dotted line); $b=30$\,cm, $U_f=40$\,cm/s, $U=3.9.$ \dotfill 26

Fig.4: Moving-film photographs of flames propagating in $10\%$ methane-air mixture in tubes of $2.5$\,cm (left) and $20$\,cm (right) diameter. Both flames travel $26$\,cm distance. {\it Source}: Reproduced from Plate I of \cite{coward1932I} \dotfill 27

Fig.5: Series of snap-shot photographs of the early stage of flame evolution after ignition, made at equal time intervals in a glass tube of $d=10$\,cm. {\it Source}: Reproduced from Figure~10 of \cite{coward1932I} \dotfill 27

Fig.6: Schematics of steady flame propagation in horizontal channel. Horizontal arrows of different length qualitatively depict gas velocity profiles in various parts of the channel. Also shown are a pair of neighboring streamlines, and transition domain indented by two vertical broken lines -- the region where $w\ne 0.$ \dotfill 27

Fig.7: Identification of physical solutions in the case $\theta = 7,$ $g=61.3$ (natural units). Solid curves represent $q_T(U)$ for Type I (left) and Type II (right) solutions, dotted curve is the plot of $(\theta^2 + 1)U^3/4.$ \dotfill 28

Fig.8: Determination of the relative error in $U$ for Type I solution with $\theta=5,$ $b=40$\,cm, $U_f = 10$\,cm/s, and $U=11.3.$ Solid curve is the plot of $w/u,$ dotted curve -- $f/U.$ The ordinate of their intersection $\approx 0.11$ \dotfill 28

Fig.9: Collection of experimental data for the speed of uniform movement (S.U.M) in tubes of various diameters. {\it Source}: Reproduced from page 1999 of \cite{coward1932I} \dotfill 29

Fig.10: Comparison of Type I (solid line) and Type II (dotted line) numerical solutions  with the experimental data for flame speed in $6\%$ (a) and $7\%$ (b) methane-air mixtures. Theoretical curves are corrected by the factor $(1-\delta)$ accounting for the heat losses. Flame speed is measured in cm/s, tube diameter -- in cm.  \dotfill 30

Fig.11: Same for $9\%$ (a) and $10\%$ (b) methane-air mixtures.\dotfill 30

Fig.12: Same for $8\%$ methane-air mixture.\dotfill 31

Fig.13: Same for $11\%$ (a), $12\%$ (b), $13\%$ (c), and $14\%$ (d) methane-air mixtures.\dotfill 32

~\newpage

\end{document}